\documentclass[11pt]{article}

\usepackage{epsf,latexsym,amssymb,amsmath,amsthm,graphics,graphicx}
\usepackage[mathscr]{eucal}

\usepackage{times}

\newtheorem{theorem}{Theorem}
\newtheorem{itlemma}{Lemma}[section]
\newtheorem{itproposition}[itlemma]{Proposition}
\newtheorem{itcorollary}[itlemma]{Corollary}
\newtheorem{itremark}[itlemma]{Remark}
\newtheorem{itremarks}[itlemma]{Remarks}
\newtheorem{itdefinition}[itlemma]{Definition}
\newtheorem{itexample}[itlemma]{Example}
\newtheorem{factenv}{Fact}

\newenvironment{corollary}{\begin{itcorollary}\rm}{\end{itcorollary}} 
\newenvironment{definition}{\begin{itdefinition}\rm}{\end{itdefinition}}
\newenvironment{example}{\begin{itexample}\rm}{\end{itexample}}
\newenvironment{lemma}{\begin{itlemma}\rm}{\end{itlemma}} 
\newenvironment{proposition}{\begin{itproposition}\rm}{\end{itproposition}}
\newenvironment{remark}{\begin{itremark}\rm}{\end{itremark}} 
\newenvironment{remarks}{\begin{itremarks} \rm}{\end{itremarks}}
\newenvironment{theo}{\begin{theorem} \rm}{\end{theorem}} 
\newenvironment{fact1}{\begin{factenv}\rm}{\end{factenv}}

\newcommand{\bc}[1]{\begin{corollary}\label{#1}}
\newcommand{\ec}{\end{corollary}}
\newcommand{\bd}[1]{\begin{definition}\label{#1}}
\newcommand{\ed}{\end{definition}}
\newcommand{\beqn}[1]{\begin{eqnarray}\label{#1}}
\newcommand{\eeqn}{\end{eqnarray}}
\newcommand{\beq}{\begin{eqnarray*}}
\newcommand{\eeq}{\end{eqnarray*}}
\newcommand{\bex}[1]{\begin{example}\label{#1}}
\newcommand{\eex}{\end{example}}
\newcommand{\bit}{\begin{itemize}}
\newcommand{\eit}{\end{itemize}}
\newcommand{\benu}{\begin{enumerate}}
\newcommand{\eenu}{\end{enumerate}}
\newcommand{\bl}[1]{\begin{lemma}\label{#1}}
\newcommand{\el}{\end{lemma}}
\newcommand{\bpr}{\begin{proof}}
\newcommand{\epr}{\end{proof}}
\newcommand{\bp}[1]{\begin{proposition}\label{#1}}
\newcommand{\ep}{\end{proposition}}

\newcommand{\rf}[1]{~(\ref{#1})}
\newcommand{\br}[1]{\begin{remark}\label{#1}}
\newcommand{\er}{\end{remark}}
\newcommand{\brs}[1]{\begin{remarks}\label{#1}}
\newcommand{\ers}{\end{remarks}}
\newcommand{\bt}[1]{\begin{theo}\label{#1}}
\newcommand{\et}{\end{theo}}
\newcommand{\bfc}[1]{\begin{fact1}\label{#1}}
\newcommand{\efc}{\end{fact1}}

\newcommand{\comment}[1]{}

\newcommand{\eps}{\varepsilon}

\newcommand{\N}{{\mathbb N}}  

\renewcommand{\S}{{\mathcal S}}

\newcommand{\tL}{\mbox{\tiny $L$}}
\newcommand{\Prot}{\mbox{\it Prot}}
\newcommand{\mRNA}{\mbox{\it mRNA}}
\newcommand{\tProt}{\mbox{{\tiny\it Prot}}}
\newcommand{\tmRNA}{\mbox{{\tiny\it mRNA}}}

\newcommand{\titleref}[1]{#1}
\newcommand{\pr}[1]{#1}
\newcommand{\nr}[1]{#1}
\newcommand{\ea}{et al.}

\begin{document}

\title{Robustness and fragility of Boolean models\\
 for genetic regulatory networks}


\author{Madalena Chaves$^{1\,*}$, R\'eka Albert$^{2}$ \& Eduardo D.\ Sontag$^{1}$}

\date{}

\maketitle

\begin{center}
{\small

$^{1}$ Department of Mathematics and BioMaPS Institute for Quantitative Biology, \\
Rutgers University, Piscataway, NJ 08854 

$^{2}$ Department of Physics and Huck Institutes for the Life Sciences, \\
Pennsylvania State University, University Park, PA 16802.

$^*$ Corresponding author, madalena@math.rutgers.edu
}
\end{center}

\begin{abstract}
Interactions between genes and gene products give rise to complex circuits
that enable cells to process information and respond to external signals. Theoretical
studies often describe these interactions using continuous, stochastic, or 
logical approaches. We propose a new modeling framework for gene regulatory networks,
that combines the intuitive appeal of a qualitative description of gene
states with a high flexibility in incorporating stochasticity in the duration 
of cellular processes. 
We apply our methods to the regulatory network of the segment polarity
genes, thus gaining novel insights into the development of gene expression patterns.  
For example, we show that very short synthesis and decay times can perturb the
wild type pattern. On the other hand, separation of timescales between pre- and
posttranslational processes and a minimal prepattern ensure convergence
to the wild type expression pattern regardless of fluctuations.
\end{abstract}

\centerline{
}

\section{Introduction}

Understanding how genetic information is translated into proteins to produce
various cell types remains a major challenge in contemporary
biology~\cite{w98}. Gene products often regulate the synthesis of mRNAs and
proteins, forming complex networks of regulatory interactions. 
Concurrently with experimental progress in gene control networks~\cite{d02}, 
several alternative modeling frameworks have been proposed. 
In the continuous-state approach, the concentrations of cellular components are
assumed to be continuous functions of time, governed by differential equations
with mass-action (or more general)
kinetics~\cite{rs95,dmmo00,grs01}. Stochastic models address the deviations
from population homogeneity by transforming reaction rates into probabilities
and concentrations into numbers of molecules~\cite{rwa02}. Finally, in the
discrete approach, each component is assumed to have a small number of
qualitative states, and the regulatory interactions are described by logical
functions~\cite{mta99,st01,ybd01,kpst03,gt03,b97,ao03}.

The kinetic details of protein-protein or protein-DNA interactions are rarely known, but there 
is increasing evidence that the input-output curves of regulatory relationships
are strongly sigmoidal and can be well approximated by step functions~\cite{ybd01,t73}. Moreover,
both models and experiments suggest that regulatory networks are remarkably robust, that is, 
they maintain their function even when faced with fluctuations
in components and reaction rates~\cite{dmmo00,asml99,eb02,cd02,cw04}. 
These observations lend support to the assumption of discrete states for genetic network 
components and of combinatorial rules for the effects of transcription factors~\cite{gk73,jg04}.
The extreme of discretization, Boolean models, consider only two states (expressed or not), 
closely mimicking the 
inference methods used in genetics~\cite{kpst03,t73,k93}. It is straightforward to study the effect 
of knock-out mutations or changes in initial conditions in this
framework, and the agreement between a real system and a Boolean model of it is a strong 
indication of the robustness of the system to changes in kinetic details~\cite{ao03}.
 
In discrete models the decision whether a network node (component) will be affected by a synthesis 
or decay process is determined by the state of effector nodes
(nodes that interact with it). Typical time-dependent Boolean models use synchronous updating 
rules~\cite{kpst03,ao03,b97,k93}, assuming that the time scales of the processes taking place in 
the system are similar.  In reality the timescales of transcription, translation, and degradation
can vary widely from gene to gene and can be anywhere from minutes to hours. Logical models following the formalism
introduced by Ren\'e Thomas~\cite{t73} allow asynchronism by associating two variables
to each gene: a state variable describing the level of its protein, 
and an image variable that is the output of the logical rule whose inputs are the state
variables of effector nodes. Whether the future state variable of a gene equals the
image or current state variable depends on the update order and, in the absence 
of temporal information, the Thomas formalism focuses
on determining the steady states, where the state and image variables 
coincide~\cite{mta99,st01,gt03,bcrg04}. The effect of asynchronous updates on the dynamics
of the system, however, has not been explored yet. 

In this paper, we present a methodology for testing the robustness of Boolean models with
respect to stochasticity in the order of updates. 
Through this, we are also probing the system itself: will individual variations
lead to unexpected gene expression patterns?
In the asynchronous method, the synthesis/decay decision is made at different time-points for each node, 
allowing individual variability in each process' duration, but more importantly, it allows for
decision reversal if the dynamics of effector nodes changes. 
It becomes possible to reproduce, e.g., the overturning of mRNA decay when its transcriptional
activator is synthesized, a process that synchronous update cannot capture. Thus,
replacing synchronous with asynchronous updates is not merely a technical detail, but rather a fundamental paradigm
shift from pointwise in time to potentially continuous communication between nodes.
Indeed, the effective synthesis or decay time for a certain node are determined by the time interval between 
the latest update of its effector nodes and its current update time, and can be any positive fraction
of the unit time interval. 
We propose three algorithms, with varying freedom in the relative duration of cellular processes, and find that 
very short transcription or decay times have the potential to derail the wild type development process.

The steady states of a Boolean model will remain the same regardless of the mechanism of 
update, but its dynamical behavior can be drastically
altered due to the stochastic nature of the updates; for instance, the same initial state may
lead to different steady states or limit cycles. Since the duration of synthesis and decay
processes is not known, we randomly explore the space of all possible timescales and update
orders, and derive the probability of different outcomes. Our methods offer a systematic way of 
exploring generic behavior of gene regulatory networks and comparing it to experimentally observed outcomes. 
To present a concrete example, we generalize a previously introduced Boolean model of the {\it Drosophila} 
segment polarity genes~\cite{ao03}. This model reproduces the wild type steady state pattern
of the segment polarity genes as well as the gene patterns of mutants, but its dynamic behavior is not directly
comparable to that of the real system. Here we show that asynchronous update leads to a much more realistic
model that gives further insights into the robustness of the gene regulatory network.

\section{The segment polarity gene network in {\it Drosophila}}

\begin{figure}
\centerline{
\scalebox{0.5}[0.5]{\includegraphics{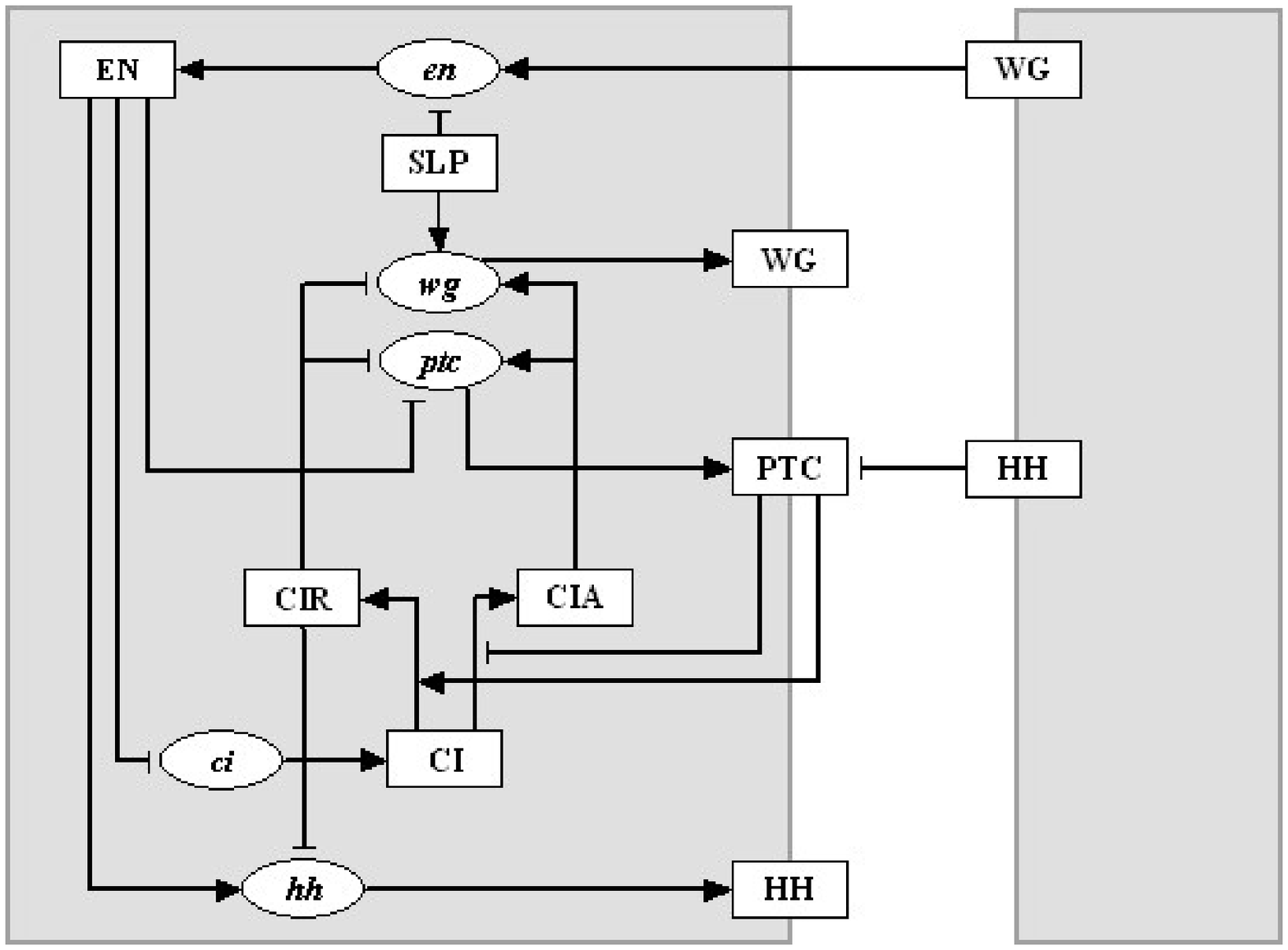} }
}
\caption{The network of interactions between the segment polarity 
genes. The grey background layers illustrate two neighboring cells,
indicating that some interactions in this network are inter-cellular.
The shape of the nodes indicates whether the corresponding
substances are mRNAs (ellipses) or proteins (rectangles). The edges of the 
network signify either biochemical 
reactions (e.g. translation,protein interactions) or 
regulatory interactions (e.g. transcriptional activation).
The edges are classified as activating ($\rightarrow$)
or inhibiting ($\dashv$ ). Figure adapted from \cite{ao03}.}
\label{fig-network}
\end{figure}

The {\it Drosophila melanogaster} segment polarity genes represent the last step in the 
hierarchical cascade of gene families initiating the segmented body of the fruit 
fly. While the preceding genes act transiently, 
the segment polarity genes are expressed throughout the life of the fly, 
and their periodic spatial pattern 
is maintained for at least $3$ hours of embryonic development~\cite{w98}. The regulatory roles
of the previously expressed genes such as the pair-rule genes {\it fushi tarazu, runt, even-skipped}
are incorporated in the prepattern (initial state) of the segment polarity genes.
The stable maintenance of the segment polarity gene expression is due to the 
interactions between these genes (see Figure \ref{fig-network}), and it is a crucial requirement in 
the development and stability of the parasegmental furrows. The best characterized segment 
polarity genes include {\it engrailed} ($en$), {\it wingless} ($wg$), {\it hedgehog} ($hh$), 
{\it patched} ($ptc$), {\it cubitus interruptus} ($ci$) and 
{\it sloppy paired} ($slp$), encoding for diverse proteins including 
transcription factors as well as secreted 
and receptor proteins.

The pair-rule gene {\it sloppy paired} ($slp$) is activated before
the segment polarity genes and expressed constitutively thereafter
 \cite{gpg92,cgg94}. {\it slp} encodes two
forkhead domain transcription factors with similar functions that
activate $wg$ transcription and repress $en$ transcription, and since
they are co-expressed we designate them both SLP. The $wg$ gene
encodes a glycoprotein that is secreted from the cells that synthesize
it \cite{hs92,pv99}, and can bind to
the Frizzled receptor on neighboring cells, initiating a signaling cascade leading to
the transcription of {\it engrailed} ($en$) \cite{cn97}. 
EN, the homeodomain-containing product of the {\it en} gene,
promotes the transcription of the {\it hedgehog} gene ({\it hh})
\cite{tek92}. In addition to the homeodomain, EN
contains a separate repression domain that
affects the transcription of $ci$ \cite{ek90} and
possibly $ptc$ \cite{hi90,tnmi93}. The
hedgehog protein (HH) is tethered to the cell membrane by a
cholesterol linkage that is severed by the dispatched protein, 
freeing it to bind to the HH receptor PTC on a neighboring cell \cite{im01}
. The intracellular domain of PTC forms a
complex with smoothened (SMO) in which
SMO is inactivated by a post-translational conformation change (Ingham
1998).  Binding of HH to PTC removes the inhibition of SMO, and
activates a pathway that results in the modification of CI \cite{i98}. The CI protein
can be converted into one of two transcription factors, depending on the PTC-HH
interactions. In the absence of HH signaling CI is cleaved to form CIR,
a transcriptional repressor that represses $wg$, $ptc$ and $hh$ transcription\cite{ak99}.
 When secreted HH binds to PTC and frees SMO,
CI is converted to a transcriptional activator, CIA, that promotes the 
transcription of $wg$ and $ptc$ \cite{ak99,ok98}.

\begin{figure}
\vskip-1cm
\centering
\includegraphics[width=13cm,angle=-90,clip=]{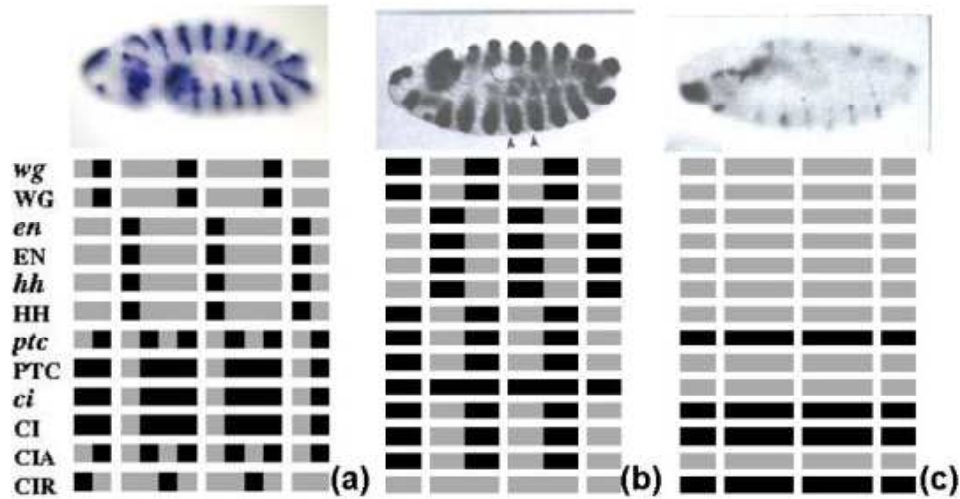}
\caption{a) Top: Illustration of the gene expression pattern of {\it wingless} on a gastrulating 
(stage 9) embryo. Other segment polarity genes have similar periodic patterns that are 
maintained for around three hours of embryonic development. The parasegmental furrows
form at the posterior border of the $wg$-expressing cells~\cite{w98}. Bottom: Synthesis of the wild
type expression patterns of the segment polarity genes (see also text)~\cite{hs92,w98}. Left corresponds to 
anterior and right to posterior in each parasegment. Horizontal rows correspond to the pattern of
individual nodes - specified at the left side of the row - over two full and two 
partial parasegments. Each parasegment is assumed to be four cells wide. A black (gray)
box denotes a node that is (is not) expressed. b) Top: {\it wingless} expression pattern in
an {\it patched} knock-out mutant embryo at stage 11~\cite{tek92}. The {\it wingless} stripes
broaden, and secondary furrows appear at the middle of the parasegment, indicating a new
{\it en-wg} boundary. Bottom: Broad striped steady state of the Boolean model, obtained
when {\it patched} is kept off (with the change that $ptc$ and PTC are not expressed), or when {\it wg, en, hh} are initiated in every cell~\cite{ao03}.
 This steady state agrees with all experimental observations on {\it ptc} mutants and
heat-shocked genes~\cite{tek92,gakt00,mbi88,slnk95,dshk88,itn91,bw93}. c) Top: {\it wingless} expression pattern in
an {\it engrailed} knock-out mutant embryo at stage 11~\cite{tek92}. The initial 
periodic pattern is disappearing, and gives rise to a non-segmented, embryonic lethal
phenotype. Bottom: Non-segmented steady state of the Boolean model, obtained when 
{\it wg, en or hh} are kept off, or cell-to-cell signaling is disrupted~\cite{ao03}. This steady state
agrees with all experimental observations on {\it wg, en, hh} mutants~\cite{tek92,dshk88,slnk95,
hi90,gakt00}. Gene expression 
images obtained from http://www.fruitfly.org (a)
and~\cite{tek92} (b,c).  }
\label{fig-drosophila}
\end{figure}

The initial state of the {\it Drosophila} segment polarity genes includes two-cell-wide SLP stripes 
followed by two-cell-wide stripes not expressing SLP \cite{cgg94}, single-cell-wide $wg$, $en$ and $hh$ stripes 
followed by three cells not expressing them, and three-cell-wide stripes for
$ci$ and $ptc$ \cite{hs92,w98}:
This pattern is maintained almost unmodified for three hours\footnote{A notable exception includes the refinement
of the $ptc$ pattern.} (see Fig.~\ref{fig-drosophila}a), 
during which time the embryo is divided into $14$ parasegments by furrows positioned between the the $wg$ and 
$en$ -expressing cells~\cite{hs92}. 

The first model of the segment polarity gene network was proposed
by von Dassow and collaborators \cite{dmmo00}, and is a continuous-state model of $13$ equations
and $48$ unknown kinetic parameters. The main conclusion of the \cite{dmmo00} article
is that the gene patterns are robust with respect to variations in the kinetic
constants in the rate laws, thus the essential feature of this network is its topology,
i.e. the existence and signature (activating or inhibiting) of the interactions. The idea of 
the network topology determining its dynamics was further explored by \cite{ao03}, who
used a slightly different network reconstruction and assumed synchronous Boolean regulation
among nodes. In the \cite{ao03} model each mRNA and protein is represented by a node of a
network, and the state of each node is $1$ or $0$, 
according to whether the corresponding substance
is present or not. The states of the nodes are updated synchronously, and the future
state of node $i$ is determined by a Boolean function of its current state and 
the current states of those nodes that
have edges incident on it. The updating functions 
are based on the experimental information and on the following
dynamical assumptions: (i) the synthesis of mRNAs/proteins has the duration of one timestep;
(ii) the effect of transcriptional activators and inhibitors is
never additive, but rather, inhibitors are dominant; 
(iii) mRNAs decay in one timestep if not transcribed;
(iv) transcription factors and proteins undergoing post-translational
modification decay in one timestep if their mRNA is not present;
(v) protein-protein binding, such as in the formation of the 
Patched-Hedgehog complex, is assumed to be instantaneous.
In summary, the \cite{ao03} model assumes that gene transcription,
protein translation, mRNA and protein decay all happen on a similar
timescale, while protein complex formation is instantaneous compared to
this common timescale.

The \cite{dmmo00} and \cite{ao03} models agree in their conclusions regarding the robustness
of the segment polarity gene network. The simplicity of the Boolean rules in the latter also allows for
the exploration of knock-out mutations and changes in the prepattern of the segment polarity genes.
Starting from the known initial state of 
{\it en, wg, hh, ptc, ci} and SLP, and assuming the null(off) state for all other nodes
the \cite{ao03} model leads to a time-invariant spatial pattern (see Fig. 2a)
that coincides with the experimentally
observed wild-type expression of the segment polarity genes
during stages $9$-$11$. Indeed, $wg$ and WG are expressed in the most posterior cell
of each parasegment, while  $en$, EN, $hh$ and HH are expressed
in the most anterior cell of each parasegment, as is observed experimentally
\cite{i98,tek92}, $ptc$ is expressed in two 
stripes of cells, one stripe on each side
of the $en$-expressing cells, the anterior one coinciding with the $wg$ stripe
\cite{hi90,hs92}. $ci$ is expressed almost ubiquitously, with the
exception of the cells expressing $en$ \cite{ek90}. CIA is  expressed in the neighbors of the HH-expressing cells,
while CIR is expressed far from the HH-expressing cells \cite{ak99}. The model indicates that 
knock-out mutations in {\it en, wg, hh} cause the non-segmented
gene pattern shown on Fig. 2b, which agrees with experimental observations.
 Indeed, the $hh$ expression in $en$ null embryos starts normally, but disappears before stage $10$
\cite{tek92}. In $wg$ null embryos, $en$ is initiated normally but
fades away by stage $9$, as observed by DiNardo {\it et al.} (1988),
 while $ci$ is ubiquitously expressed \cite{slnk95}. In $hh$ mutant embryos 
 the $wg$ expression disappears by stage $10$ \cite{hi90}, as does the expression of $ptc$,
 and there is no segmentation \cite{gakt00}. On the other hand, {\it ptc} knockout mutations or
overexpressed initial states lead to the broad-striped pattern of Fig. 2c\footnote{The only difference between the
{\it ptc} mutant and heat-shock pattern is that the former does not express $ptc$ and PTC}. Indeed, 
experimental results indicate broad $en$, $wg$ and $hh$ stripes 
 \cite{tek92,gakt00,mbi88} and  Gallet {\it et al.} (2000) 
 find that a new ectopic groove forms at the second $en-wg$ interface at the middle of the
parasegment. Also, $ci$ is not expressed at this ectopic groove \cite{slnk95}.
 In heat-shock experiments the $wg$ and $ptc$ 
stripes expand anteriorly when $hh$ or $en$ are
ubiquitously induced \cite{gakt00}, while
narrower $ci$ stripes emerge after a transient decay of $ci$ \cite{slnk95}.
Intriguingly, the \cite{ao03} model finds that a knock-out mutation of $ci$
does not change the {\it en, wg, hh} patterns but disrupts $ptc$ expression;
experiments indicate that the segmental
grooves are present and $wg$ is expressed until stage $11$, but $ptc$
expression decays \cite{gakt00}. In summary, the simple synchronous Boolean 
model~\cite{ao03} 
captures perfectly the wild type and mutant expression patterns of the segment
polarity genes, and thus serves as a good starting point for a more realistic 
model that relaxes the assumption of synchronicity.

We focus our attention on a single parasegment of four cells, thus the total
number of nodes we consider is $4\times 13=52$. We use the same interaction topology and logical rules as the 
synchronous model \cite{ao03}, but instead of assuming that the states of all nodes are updated simultaneously, 
we update the state of each node individually (see Table~\ref{table_rules}). To maintain the highest generality, 
we incorporate possible cell to cell variations in synthesis and decay processes.\footnote{We follow the 
\cite{ao03} model in assuming very short 
timescales for PTC-HH binding and
SMO activation, and consequently in Figure \ref{fig-network} and in the regulatory rules we connect the CI posttranslational modifications
to HH signaling. We have verified that this assumption can be relaxed without any qualitative changes in
the results.}  
\begin{table}
\caption{Regulatory functions governing the states of segment polarity gene products in the model
. Each node is labeled by its biochemical symbol and subscripts signify cell number. The times
 $\tau^j$ signify the last time node $j$ was updated before $t$.}
\label{table_rules}
\centering{
\begin{tabular}{ll}
\hline 
Node & Boolean updating function in the asynchronous algorithm\\
\hline\hline
$SLP_i$ & $SLP_i(t)=\left\{\begin{array}{lllll}
0 &\mbox{if}& i\in\{1,2\}\\
1 &\mbox{if}& i\in \{3,4\}\\
\end{array}\right.$ \\
$wg_i$ & $wg_i(t)=(CIA_i(\tau_{CIA})$ and $SLP_i(\tau_{SLP})$ and not $CIR_i(\tau_{CIR}))$\\
        & or $[wg_i(\tau_{wg}) $ and $(CIA_i(\tau_{CIA})$ or $SLP_i(\tau_{SLP}) )$ 
		 and not $CIR_i(\tau_{CIR})]$ \\
$WG_i$ & $WG_i(t)=wg_i(\tau_{wg})$  \\
$en_i$ & $en_i(t)=(WG_{i-1}(\tau_{WG1})$ or $WG_{i+1}(\tau_{WG2}))$ and not $SLP_i(\tau_{SLP})$ \\
$EN_i$ & $EN_i(t)=en_i(\tau_{en})$  \\
$hh_i$ & $hh_i(t)=EN_i(\tau_{EN})$ and not $CIR_i(\tau_{CIR})$ \\
$HH_i$ & $HH_i(t)=hh_i(\tau_{hh})$ \\
$ptc_i$ & $ptc_i(t)=CIA_i(\tau_{CIA})$ and  not $EN_i(\tau_{EN})$ and not $CIR_i(\tau_{CIR})$ \\
$PTC_i$ & $PTC_i(t)=ptc_i(\tau_{ptc})$ or $(PTC_i(\tau_{PTC})$ 
and not $HH_{i-1}(\tau_{HH1})$ and not $HH_{i+1}(\tau_{HH2}))$ \\
$ci_i$ & $ci_i(t)=$ not $EN_i(\tau_{EN})$ \\
$CI_i$ & $CI_i(t)=ci_i(\tau_{ci})$ \\
$CIA_i$ & $CIA_i(t)=CI_i(\tau_{CI})$ and [not $PTC_i(\tau_{PTC})$ or $HH_{i-1}(\tau_{HH1})$\\ 
        &   or $HH_{i+1}(\tau_{HH2})$ or $hh_{i-1}(\tau_{hh1})$ or $hh_{i+1}(\tau_{hh2})$]\\
$CIR_i$ &  $CIR_i(t)=CI_i(\tau_{CI})$ and $PTC_i(\tau_{PTC})$ 
         and not $HH_{i-1}(\tau_{HH1})$ and not $HH_{i+1}(\tau_{HH2})$\\
       &  and not $hh_{i-1}(\tau_{hh1})$ and not $hh_{i+1}(\tau_{hh2})$\\
\hline
\end{tabular}}
\end{table}

Throught the text, the notation ``$wg_1^t$'' or ``$wg_1(t)$'' represent the state of 
{\it wingless} mRNA in the first cell of 
the parasegment at time $t$. Similar notations apply for other mRNAs and proteins. There 
are 4 cells in 
each parasegment, and  we adopted periodic boundary conditions, meaning that: $node_{4+1}=node_1$ and 
$node_{1-1}=node_4$. The wild type initial state corresponds to: 
\beqn{eq-initial-wt}
   wg_4^0=1,\ \ en_1^0=1,\ \ hh_1^0=1,\ \ ptc_{2,3,4}^0=1, \ \ ci_{2,3,4}^0=1
\eeqn
and the remaining nodes are zero.
The asynchronous model represented in Table~\ref{table_rules} exhibits the same steady states as the 
synchronous model developed in~\cite{ao03}. Note that three of the four main steady states
agree perfectly with experimentally observed states corresponding to wild type, 
{\it en, wg or hh} mutant and {\it ptc} mutant embryonic patterns~\cite{tek92,dshk88,slnk95,
hi90,gakt00,mbi88,bw93,itn91,hs92,w98}. A summary is presented in Table~\ref{table_steady_states}.
\begin{table}[h]
\label{table_steady_states}
\caption{Complete characterization of the model's steady states.}
\begin{center}
\begin{tabular}{ll}
\hline
  Steady state  & Expressed nodes \\
\hline\hline
  wild type & $wg_{4}$, $WG_{4}$, $en_{1}$, $EN_{1}$, $hh_{1}$, $HH_{1}$, \\
            & $ptc_{2,4}$, $PTC_{2,3,4}$, $ci_{2,3,4}$, 
              $CI_{2,3,4}$, $CIA_{2,4}$, $CIR_{3}$   \\
\hline
  broad stripes & $wg_{3,4}$, $WG_{3,4}$, $en_{1,2}$, $EN_{1,2}$, 
                       $hh_{1,2}$, $HH_{1,2}$,  \\
                & $ptc_{3,4}$, $PTC_{3,4}$, $ci_{3,4}$, $CI_{3,4}$, $CIA_{3,4}$  \\
\hline
  no segmentation & $ci_{1,2,3,4}$, $CI_{1,2,3,4}$, 
                    $PTC_{1,2,3,4}$, $CIR_{1,2,3,4}$ \\
\hline
  wild type variant & $wg_{4}$, $WG_{4}$, $en_{1}$, $EN_{1}$, $hh_{1}$, $HH_{1}$, \\
            & $ptc_{2,4}$, $PTC_{1,2,3,4}$, $ci_{2,3,4}$, 
              $CI_{2,3,4}$, $CIA_{2,4}$, $CIR_{3}$ \\
\hline
  ectopic  &  $wg_{3}$, $WG_{3}$, $en_{2}$, $EN_{2}$, $hh_{2}$, $HH_{2}$, \\
            & $ptc_{1,3}$, $PTC_{1,3,4}$, $ci_{1,3,4}$, 
              $CI_{1,3,4}$, $CIA_{1,3}$, $CIR_{4}$   \\
\hline
  ectopic variant & $wg_{3}$, $WG_{3}$, $en_{2}$, $EN_{2}$, $hh_{2}$, $HH_{2}$, \\
            & $ptc_{1,3}$, $PTC_{1,2,3,4}$, $ci_{1,3,4}$, 
              $CI_{1,3,4}$, $CIA_{1,3}$, $CIR_{4}$  \\
\hline
\end{tabular}
\end{center}
\end{table}

\section{Randomly perturbed timescales}

As in the context of parallel computation systems, the fundamental difference between 
synchronous and asynchronous updates is at the level of task coordination and data communication  
among nodes in a network~\cite{tsi}.
Synchronous algorithms are highly coordinated: at pre-determined instants, all the nodes  
``stop'' and exchange the current information among themselves.
For instance, suppose there are $N$ nodes, where each node $i$ 
``computes'' the state of variable $x_i$, according to a function $f_i(x_1,x_2,\ldots,x_N)$ 
($i=1,\ldots,N$). When all the $N$ nodes have finished phase $k$, they exchange
their current states, $x_i^k$, and then proceed to phase $k+1$, that is
\beq
   x_i^{k+1}=f_i(x_1^k,x_2^k,\ldots,x_N^k).
\eeq
Asynchronous algorithms, on the other hand, admit a greater flexibility at the level of process coordination. 
Each node is allowed to have its own ``computation rate'', that is, during any time interval $[t_a,t_b]$,
node $i$ may be updated only once, while node $j$ may be updated $\ell>1$ times. 
In this case, communication delays between nodes may occur, and some possibly outdated 
information may be used: for instance, node $j$ uses the same value $x_i(t_a)$ throughout its $\ell$ updates in
the interval $[t_a,t_b]$. However, an overall gain in efficiency in achieving the final result may be expected.
For instance, in our example, the wild type steady state is reached in less than 4 steps with the 
asynchronous algorithms (see Sections~\ref{sec-timesep},~\ref{sec-markov}), 
while with the synchronous algorithm 6 steps are needed~\cite{ao03}.

In general, we may say that node $i$ updates its state at times:
\beq
    T_i^1, T_i^2,\ldots, T_i^k,\ldots \ \ \  k\in\N_0,
\eeq
and the local variables, $x_i$, are updated according to:
\beqn{eq-inode}
      x_i[T_i^k] = f_i(x_1[\tau^k_{1i}],\ldots,x_N[\tau^k_{Ni}]),
\eeqn
where $\tau^k_{ji}$ is defined as
\beq
      \tau^k_{ji}=\mbox{ the latest available communication to node $i$, from node $j$.}
\eeq
There is usually a distinction between {\em totally} or {\em partially} asynchronous algorithms: 
the latter impose an updating constraint (every variable is updated at least once in any interval of a 
fixed length), while the former simply ensure that a variable is updated infinitely many times.

In a first numerical experiment we consider a totally asynchronous algorithm, with the highest degree of 
individual variability in each process' duration. The time unit of the synchronous model is randomly perturbed, 
so that the set of updating times for each node $i$ ($1\leq i\leq N$) is of the form 
\beq
   T_i^{k+1} = T_i^k +1 +\eps\;r_i^k,\ \ \  k\in\N,
\eeq
where $r_i^k$ are random numbers generated at each iteration, out of a uniform distribution in the interval 
$[-1,1]$. The value $\eps\in[0,1)$ is the magnitude of the
perturbation (the case $\eps=0$ coincides with the synchronous algorithm). At any given time $t$, the next node(s) 
to be updated is(are) $j$ such that $T_j^{\ell}=\min_{i,k}\{ T_i^k\geq t\}$, for some $\ell$.
Since the duration of synthesis and decay processes is not known, through this algorithm one may randomly explore the 
space of all possible timescales and update orders, and derive the probability of different outcomes. 
The set of updating times $\{T_i^k, k\in\N_0 \}$ may vary with each execution of the algorithm, so an element of
stochasticity is naturally introduced.

\begin{figure}
\label{fig-allstates}
\centerline{
\scalebox{0.55}[0.55]{\includegraphics{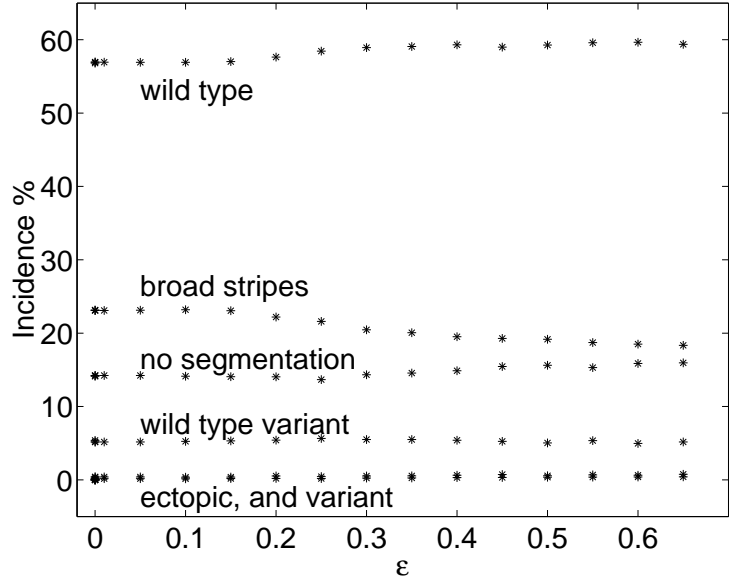} } }
\caption{Fragility of the regulatory network. With the totally asynchronous algorithm 
the wild type initial state can lead to one of the six distinct steady states. 
Each $*$ corresponds to an $\eps$ perturbation of the unit time-step. 
Note that the $\eps \rightarrow 0$ limit does not give the same results as a synchronous update,
demonstrating the fundamental difference between synchronous and asynchronous models.}
\end{figure}

Always starting from the wild type initial condition\rf{eq-initial-wt}, this experiment was conducted over a 
wide range of perturbations ($10^{-12}\leq\eps\leq0.65$), and 30000 trials were executed for each $\eps$. 
The results (see Figure~\ref{fig-allstates}) show that all of the model's steady states may occur with a certain
frequency: the wild type pattern with only $57\%$, followed by the broad-striped pattern ($24\%$) observed in 
heat-shock experiments and $ptc$ mutants~\cite{gakt00} and by the pattern with no segmentation ($15\%$) observed 
in $en$, $hh$ or $wg$ mutants~\cite{tek92}, the latter two corresponding to embryonic lethal phenotypes~\cite{gakt00}. 

We observe that each of the steady state patterns {\it occurs with a frequency which is independent 
of the value of $\eps$, for $\eps<0.15$}. This may indicate that it is the order in which the protein 
and mRNA nodes are updated that determines the steady state pattern.
In order to test this hypothesis, we designed a second experiment assuming that
\bit
\item[(A1)] Every node is updated exactly once during each unit time interval $(k,k+1]$ ($k=0,1,2,\ldots$), 
    according to a given order $\phi^k$. 
\eit
This order $\phi^k$ is a permutation of $\{1,\ldots N\}$, chosen randomly (again out of a uniform 
distribution over the set of all $N!$ possible permutations) at the beginning of the time unit $k$. 
Then we have
\beq
   T_i^k = N(k-1) + \phi^k(i),\ \ \  k\in\N,
\eeq
so that $\phi^k(j)<\phi^k(i)$ implies $T_j^k<T_i^k$, and node $j$ is updated before node $i$.
The partially asynchronous algorithm leads to the same patterns, with incidence rates very similar to 
those observed with the totally asynchronous algorithm (see Table~\ref{tab-all-permutations}).
\begin{table}
\label{tab-all-permutations}
\caption{The frequencies of the six steady states observed in the partially asynchronous model confirm 
those observed for the totally asynchronous model. The frequencies are computed from $30000$ executions.}
\begin{center}
\begin{tabular}{ll}
\hline
 Steady State  & Incidence \\
\hline\hline
  wild type & 56\% \\
  broad stripes & 24\% \\
  no segmentation & 15\% \\
  wild type variant & 4.2\% \\
  ectopic  & 0.98\% \\
  ectopic variant & 0.68\% \\
\hline 
\end{tabular}
\end{center}
\end{table} 

These results indicate the fragility of the wild type gene pattern with respect to changes in the 
timescales of synthesis and decay processes. While more than half of the random timescale combinations 
still lead to the expected outcome, a considerable percentage results in loss of the prepattern and an 
inviable final state.

\subsection{Imbalance between CIA and CIR}

Further analysis shows that the divergence from wild type can be attributed to an imbalance between 
the two opposing Cubitus Interruptus transcription factors (CIA, CIR) in the posterior half of the parasegment.
Indeed, the expression of CIA and CIR in both the broad stripes and the no segmentation patterns is 
clearly distinct from that in the wild type pattern. In the next set of numerical experiments, we explore the
effects of CIA/CIR expression in the formation of the final pattern.

In wild type, the two Cubitus Interruptus proteins, CIA and CIR, are expressed in different cells of the posterior 
part of the parasegments, namely,
\beq
  & CIA_{3}=0, \ \ \ CIA_{4}=1, \\
  & CIR_{3}=1, \ \ \ CIR_{4}=0, 
\eeq
and the maintenance of these complementary ON/OFF states is essential in the wild type pattern.
To investigate the effect of an imbalance between the two Cubitus Interruptus proteins, we considered two 
disruptive cases: the (transient) overexpression of CIR, or the (transient) overexpression of CIA and absence 
of CIR in both posterior cells.

More precisely, in the totally asynchronous algorithm (choosing $\eps=0.1$), 
we transiently imposed an expression pattern for the Cubitus proteins as follows: 
\bit
\item[(a)] $CIA_{3,4}^t=1$ and $CIR_{3,4}^t=0$, for $t\in[3,3+\tau]$;
\item[(b)] $CIR_{3,4}^t=1$, for $t\in[3,3+\tau]$,
\eit
where $\tau$ is the duration of the transient. The overexpression starts after three unit time steps.
The duration of the transient was:
\beq
   \tau\in\{0,0.3,0.75,1.5,2.75,3\},
\eeq
so when $\tau=0$ the results of the general totally asynchronous algorithm are recovered.

Our results show that even a small transient imbalance between CIA and CIR causes a clear bias towards a 
mutant state: the broad stripes mutant in case (a), or the no segmentation mutant in case (b).
Thus any perturbation that leads to such an imbalance has as severe effects as 
a mutation in $ptc$ (causing the broad striped pattern) or either of $en$, $wg$ or $hh$ 
(causing the nonsegmented pattern).   

\begin{figure}[htb]
\centerline{\scalebox{0.42}[0.42]{\includegraphics{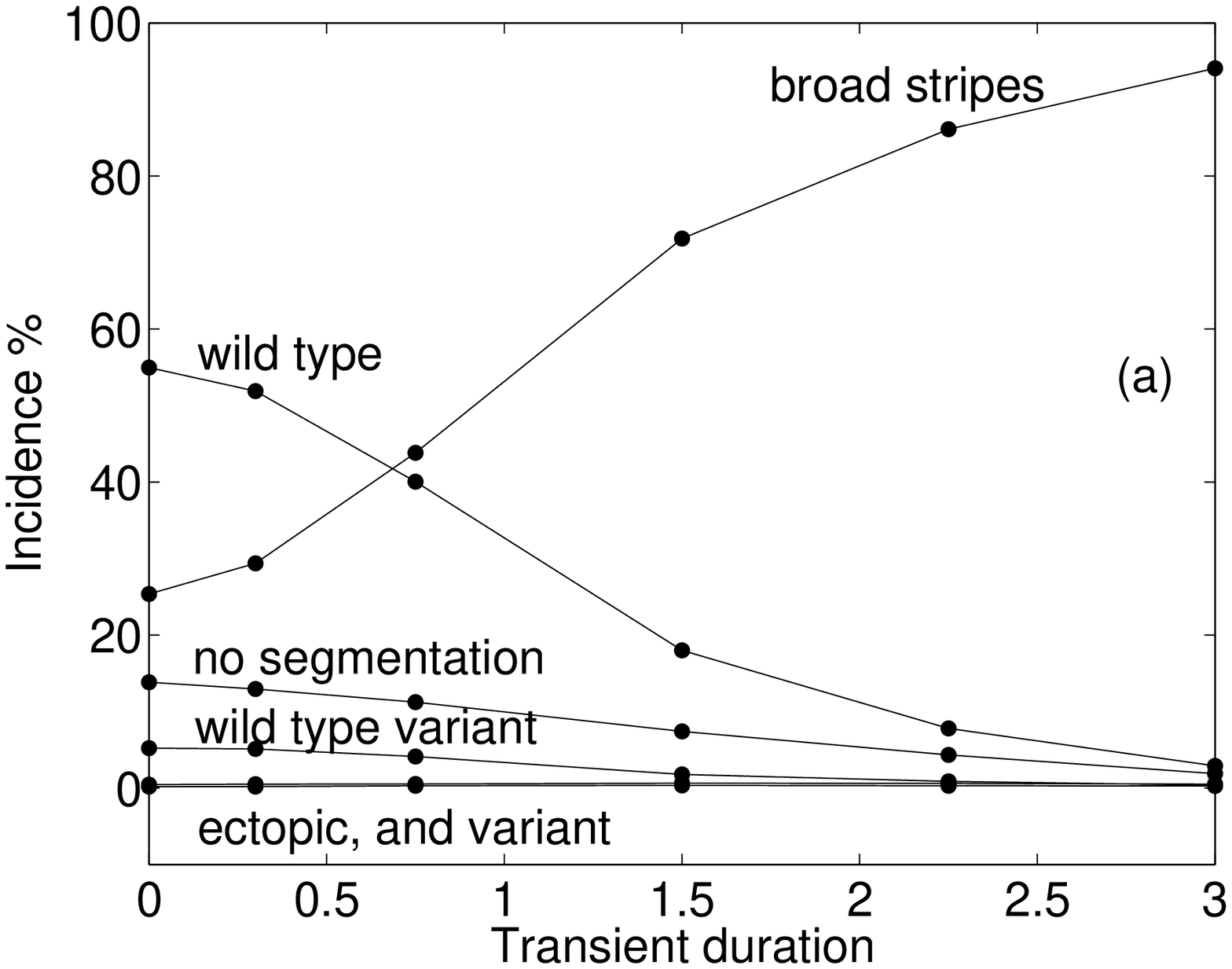}}
\scalebox{0.42}[0.42]{\includegraphics{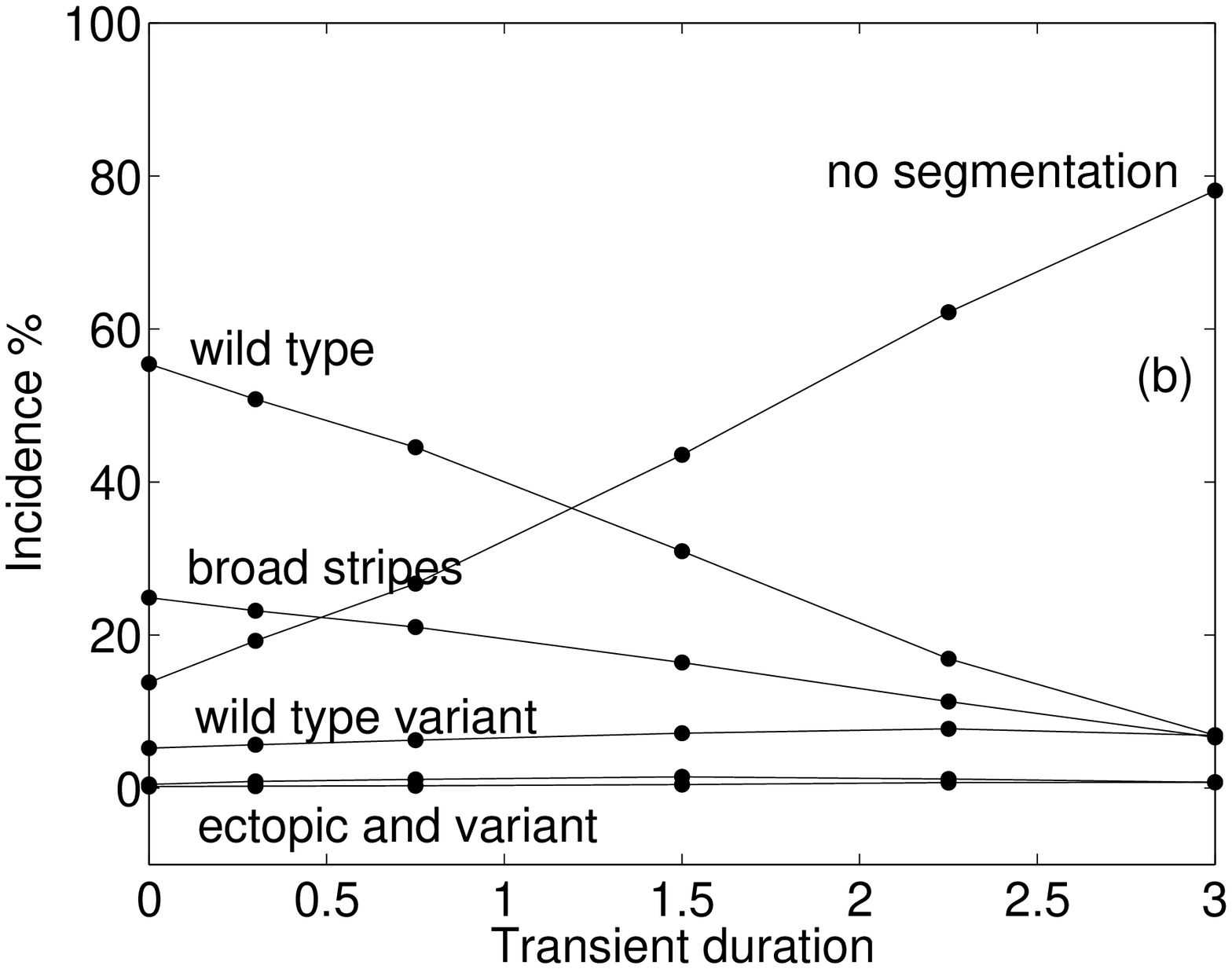}}}
\caption{Bias towards mutant states. The $x$-axis represents the duration of the 
transient, $\tau$ (in unit time steps). The incidence
probabilities were computed over 20000 trials. 
(a) The case $CIA_{3,4}^t=1$ and $CIR_{3,4}^t=0$ leads to the broad striped pattern. 
(b) The case $CIR_{3,4}^t=1$ leads to the no segmentation pattern.}
\label{fig-CIACIR}
\end{figure}

These numerical experiments also open the way to many other questions: 
are there particular sequences that lead to a given steady state? 
How is the evolution from the initial to steady state?
How robust is the asynchronous model with respect to initial conditions?

\section{Timescale separation uncovers robustness of the model}
\label{sec-timesep}
 
In both of the previous algorithms we assumed no bias towards a preferred protein/mRNA updating sequence and, 
as a result, an unrealistic divergence from the wild type pattern is observed, with high incidence of 
inviable states.
Based on the fact that post-translational processes such as protein conformational changes
or complex formation usually have shorter durations than transcription, 
translation or mRNA decay, we introduce a distinct timescale separation by choosing to update proteins 
first and mRNAs later. This leads to a model which is very robust, in the sense that the wild type pattern 
occurs with a frequency of $87.5\%$ and only one other steady state is observed, the broad striped pattern,
with a frequency of $12.5\%$.
We completely characterize this model by theoretically showing that only two of the six steady states are 
possible (and occur with well determined frequencies), and identifying the order of updates that leads to 
divergence from wild type. We also show that the wild type state is really an attractor for the system, while 
the pathway to the broad stripes state may show oscillatory cycles.

Assuming that 
\bit
\item[(A2)] All the proteins are updated before all the genes,
\eit
the $k$-th iteration of the two-timescale algorithm proceeds as follows: 
\bit
\item[(A3)] At the begining of the $k$-th time unit, generate a random permutation, $\phi^k_{\tProt}$ of 
$\{1,\ldots L\}$, and a random permutation, $\phi^k_{\tmRNA}$ of $\{L+1,\ldots N\}$ (using a uniform d
istribution over, respectively, the sets of $L!$ and $(N-L+1)!$ possible permutations).
Then the $N$ nodes are updated in the order given by $\phi^k=(\phi^k_{\tProt},\;\phi^k_{\tmRNA})$, 
according to\rf{eq-inode}, with
\beq
      \tau^k_{ji}=\left\{ \begin{array}{cr}
                            T_j^{k-1}, & \phi^k(j)\leq\phi^k(i) \\
                            T_j^k, & \phi^k(j)>\phi^k(i).
                          \end{array}
                  \right. 
\eeq
\eit
As an example, suppose that
\beq
   N=5,\ \ L=3,\ \ \phi^1_{\tProt}=\{2,1,3\}, \ \ \phi^1_{\tmRNA}=\{5,4\}.
\eeq
Then, $\phi^1=\{2,1,3,5,4\}$, and 
$T_1^1=2$, $T_2^1=1$, $T_3^1=3$, $T_4^1=5$, $T_5^1=4$.
The nodes are updated as follows
(for simplicity of notation, we will write $x_i^k:=x_i[T_i^k]$):
\beq
   x_2^1 &=& f_2(x_1^0,x_2^0,x_3^0,x_4^0,x_5^0),\\
   x_1^1 &=& f_1(x_1^0,x_2^1,x_3^0,x_4^0,x_5^0),\\
   x_3^1 &=& f_3(x_1^1,x_2^1,x_3^0,x_4^0,x_5^0),\\
   x_5^1 &=& f_5(x_1^1,x_2^1,x_3^1,x_4^0,x_5^0),\\
   x_4^1 &=& f_4(x_1^1,x_2^1,x_3^1,x_4^0,x_5^1).
\eeq

Some general inferences about the updating rules can be made. For example, the translation process only 
depends on the presence of the transcript, which is decided in the previous time unit, 
thus $\Prot^t=\mRNA^{t-1}$.
The beginning of a transcription process depends on the presence of transcription factors, and
since mRNAs are updated after proteins, $\mRNA^t=\Prot^t$. The outcome of post-translational
processes depends on the order of updates, for example the rule for a binding process will be 
$\mbox{\it Complex}^t=\Prot_1^{t_1}\mbox{ and } \Prot_2^{t_2}$, where $t_1$ and $t_2$ 
can be either $t-1$ or $t$ (see Table~\ref{table_two_rules}).

\begin{table}
\caption{Regulatory functions governing the states of segment polarity gene products in the two-timescale 
asynchronous algorithm. Each node is labeled by its biochemical symbol, subscripts signify cell number
and superscripts signify timestep. 
Although the updating time of each node varies, each function can be written by using the states of effector 
nodes at the previous or current timesteps. 
The individual times $t_1 \ldots t_{10}$ can take the values  $\{t-1, t\}$.}
\label{table_two_rules}
\centering{
\begin{tabular}{ll}
\hline 
Node & Boolean updating function in the two-timescale algorithm\\
\hline\hline
$wg_i$ & $wg_i^{t}=(CIA_i^{t}$ and $SLP_i^{t}$ and not $CIR_i^t)$
         or $[wg_i^{t-1} $ and $(CIA_i^t$ or $SLP_i^t )$ and not $CIR_i^t]$ \\
$WG_i$ & $WG_i^{t}=wg_i^{t-1}$  \\
$en_i$ & $en^{t}_i=(WG_{i-1}^t$ or $WG_{i+1}^t)$ and not $SLP^t_i$ \\
$EN_i$ & $EN^{t}_i=en^{t-1}_i$  \\
$hh_i$ & $hh_i^{t}=EN_i^t$ and not $CIR_i^t$ \\
$HH_i$ & $HH^{t}_i=hh^{t-1}_i$ \\
$ptc_i$ & $ptc_i^{t}=CIA^t_i$ and  not $EN_i^t$ and not $CIR^t_i$ \\
$PTC_i$ & $PTC^{t}_i=ptc_i^{t-1}$ or $(PTC_i^{t-1}$ and not $HH_{i-1}^{t_1}$ and not 
$HH_{i+1}^{t_2})$ \\
$ci_i$ & $ci_i^{t}=$ not $EN_i^t$ \\
$CI_i$ & $CI_i^{t}=ci_i^{t-1}$ \\
$CIA_i$ & $CIA_i^{t}=CI_i^{t_3}$ and (not $PTC_i^{t_4}$ or $HH_{i-1}^{t_5}$ or 
$HH_{i+1}^{t_6}$ or $hh_{i\pm1}^{t-1}$)\\
$CIR_i$ &  $CIR_i^{t}=CI_i^{t_7}$ and $PTC_i^{t_8}$ and not $HH_{i-1}^{t_9}$ and not $HH_{i+1}^{t_{10}}$
and not $hh_{i\pm1}^{t-1}$\\
\hline
\end{tabular}}
\end{table}

\subsection{Two steady states}

The trajectory of the system is thus defined by a sequence of permutations (obtained as described in A3) 
and the corresponding sequence of states:
\beqn{eq-syst}
     \{ \phi^k\},\ \{ x^k\}\ \ \mbox{ for } \ k=0,1,2,\ldots.
\eeqn
We will show that for pre-patterns that satisfy  $wg_4^{0}=1$, $ci_1^{0}=0$ and  $ptc_1^{0}=0$ (which 
include the pattern observed in the wild type at stage 8), the only possible steady states for 
system\rf{eq-syst} are the wild type pattern experimentally observed at stages 9-11, and a mutant 
with broad $wg$ stripes.
We assume that all the proteins are absent initially (at $T=0$), and that the which {\it sloppy pair} gene
is maintained at a constant value:  $SLP_{1,2}=0$ and $SLP_{3,4}=1$
This pattern for SLP is responsible for permanent absence (or expression) of some of the segment polarity genes, 
and corresponding proteins, in certain cells of the parasegment. By direct inspection of the model, it follows that 
\beqn{eq-wg12}
   wg_{1,2}^0=0,\ \ \Rightarrow\ \
   wg_{1,2}^T=0,\ \   WG_{1,2}^T=0,\ \ \mbox{for } T\geq0
\eeqn
\beqn{eq-en-hh34}
   en_{3,4}^T=0,\ \ EN_{3,4}^T=0,\ \ \mbox{for } T\geq0,\\
   hh_{3,4}^T=0,\ \ HH_{3,4}^T=0, \ \ \mbox{for }  T\geq0 
\eeqn
and 
\beqn{eq-ci34-T=0}
   ci_{3,4}^0=1,\ \ \Rightarrow\ \
   ci_{3,4}^T=1,\mbox{ for }  T\geq0,\ \mbox{ and }\ CI_{3,4}^T=1, \mbox{for }  T\geq1\\
   \label{eq-ci34-T=1}
   ci_{3,4}^0=0,\ \ \Rightarrow\ \
   ci_{3,4}^T=1,\mbox{ for }  T\geq1,\ \mbox{ and }\ CI_{3,4}^T=1, \mbox{for }  T\geq2.
\eeqn
The next statement reflects the fact that the effect of $wg_4$ activating $en_1$ propagates
to inhibit $ci_1$ which then eliminates all forms of CI from the first cell. 

\bfc{fc-CIR1}
Assume that $wg_4^{0}=1$, $ci_1^{0}=0$ and  $ptc_1^{0}=0$. 
For any $T\geq0$, if $wg_4^{t}=1$ for all $0\leq t\leq T$, then 
$CI_1^{t}=0$ for all $3\leq t\leq T+3$ and
$CIR_1^{t}=0$ for all $0\leq t\leq T+3$.
\efc

\bp{pr-two-stst}
Assume $wg_4^{0}=1$, $ci_1^{0}=0$ and  $ptc_1^{0}=0$. 
Under assumptions A1-A2, $wg_4^T=1$, for all $T\geq0$.
\ep

\bpr
We will argue by contradiction.
Suppose that there do exist times $t\geq1$ with $wg_4^t=0$, and
let $T$ be the minimum of such times, that is,
\beq
   wg_4^T=0\ \ \mbox{ and } wg_4^t=1,\ \ \mbox{ for all } 0\leq t<T.
\eeq 
>From the model's equations, together with assumptions A1-A2:
\beq
   WG_4^t &= & wg_4^{t-1},  \\
   wg_4^t &=& (CIA_4^t\mbox{ and not }CIR_4^t) \mbox{ or } (wg_4^{t-1}\mbox{ and not }CIR_4^t),
\eeq 
for all $t\geq1$.
So, it follows that
\beqn{eq-WG4}
   WG_4^t & =& 1,\ \ \mbox{ for all } 0\leq t\leq T, \\
   CIR_4^t & =& 0,\ \ \mbox{ for all } 0\leq t<T,
   \ \ \mbox{ and } CIR_4^T=1.
\eeqn
Now, from Fact~\ref{fc-CIR1} it also follows that
\beqn{eq-cir1}
   CIR_1^t=0 \ \ \mbox{for all}\ \ 0\leq t\leq T+2. 
\eeqn
The equation for $CIR_4$ is:
\beqn{eq-CIR4}
    CIR_4^t &=& CI_4^{t_d} \mbox{ and not } 
        [\mbox{not }PTC_4^{t_a} 
         \mbox{ or } HH_{3}^{t_c}\mbox{ or }HH_{1}^{t_b}\mbox{ or } 
                     hh_{3}^{t-1}\mbox{ or } hh_{1}^{t-1} ] \\
\eeqn
(where $t_a,\ldots, t_d\in\{ t,t-1\}$ depend on the permutation $\phi^t$). 
Recall also that
\beqn{eq-hh1}
   hh_1^{t} &=& EN_1^{t}\mbox{ and not }CIR_1^{t} \\
   \label{eq-EN1}
   EN_1^t &=& en_1^{t-1} \\
   \label{eq-en1}
   en_1^{t-1} &=& WG_4^{t-1} \mbox{ or } WG_2^{t-1}. 
\eeqn
From\rf{eq-CIR4}:
\beq
   CIR_4^T=1\ \ \Rightarrow\ \ hh_1^{T-1}=0, 
\eeq
and then from\rf{eq-cir1} and\rf{eq-hh1}:
\beq
   hh_1^{T-1}=0\ \ \Rightarrow\ \ EN_1^{T-1}=0.
\eeq
Now by equations~(\ref{eq-EN1},~\ref{eq-en1}):
\beq
   EN_1^{T-1}=0, \ \Rightarrow\ 
   en_1^{T-2}=0\  \Rightarrow\  WG_4^{T-2}=0 \mbox{ and } WG_2^{T-2}=0,
\eeq
which contradicts equation\rf{eq-WG4}.
Thus, it must be that $wg_4^T=1$ for all times $T$, as we wanted to show.   
\epr

The following are now immediate conclusions from the model.

\bc{cor-var}
$CIR_4^T=0$ for all $T\geq0$, 
$en_1^T=1$ and $WG_4^T=1$ for all $T\geq1$.
$EN_1^T=1$, $ci_1^T=0$ and $hh_1^T=1$ for all $T\geq2$.
$CI_1^T=0$ and $HH_1^T=1$ for all $T\geq3$.
And finally, $CIA_1^T=CIR_1^T=0$ for all $T\geq4$.
\ec

\bc{cor-var2}
$ptc_1^T=0$ and $PTC_1^T=0$ for all $T\geq0$, and $CIR_2^T=0$ for all $T\geq3$. 
\ec

\comment{
\bpr
The equations for these three nodes are, respectively,
\beq
   ptc_1^{T} &=& CIA_1^T\mbox{ and not } EN_1^T \mbox{ and not } CIR_1^T\\
   PTC_1^{T} &=& ptc_1^{T-1} \mbox{ or }[PTC_1^{T-1} 
                           \mbox{ and not } HH_2^{t_a} \mbox{ and not } HH_4^{t_b}] \\
   CIR_2^T &=& CI_2^{t_a} \mbox{ and }PTC_2^{t_b}\mbox{ and not }
              hh_1^{t_c}\mbox{ and not } hh_3^{t_d}\mbox{ and not }
              HH_1^{t_e}\mbox{ and not } HH_3^{t_f},
\eeq
and the initial condition satisfies $PTC_1^0=0$, $ptc_1^0=0$.
>From Corollary~\ref{cor-var}, $EN_1^T=1$ for all $T\geq2$ implies
$ptc_1^T=0$ for $T\geq2$. But, since $ci_1^0=0$ implies $CI_1^1=0$ and
$CIA_1^{0,1}=0$, also $ptc_1^1=0$.
$hh_1^T=1$ for $T\geq2$ implies $CIR_2^T=0$ for $T\geq3$.
\epr
}%

In conclusion, from Proposition~\ref{pr-two-stst} it is clear that neither the no segmentation nor the 
two ectopic patterns are steady states of the system\rf{eq-syst} under assumptions A1-A2, because all of these 
states imply $wg_4=0$. In addition, Corollary~\ref{cor-var2} shows that the wild type variant, where $PTC$ 
is ubiquitous, cannot be a steady state. Also, any of the states with $wg_{1,2}=1$ is immediately prevented by 
the initial condition\rf{eq-wg12}. This leaves only the ``regular'' wild type or the mutant with broad
$wg$ stripes.

\subsection{Divergence from wild type}

Under assumptions A1-A2, divergence from the wild type pattern occurs if and only if the first permutation 
(in particular $\phi^1_{\tProt}$) is of a particular form. Thus, convergence (or divergence) to the wild 
type pattern is decided at the first iterate ($T=1$).

Recall that the wild type pattern requires {\it wingless} not to be expressed in the third cell ($wg_3=0$). 
The next Fact (proved in the Appendix) essentially says that a stable $wg_3=0$ induces the absence of both
{\it engrailed}, {\it hedgehog} in the second cell, as well as the absence of CIA$_3$, and maintains the 
expression of $PTC_3$. 
\bfc{fc-wg3=0} 
Assume $ptc_3^0=1$ and $en_2^0=0$.
\bit
\item[(a)] Let $T\geq1$. If $wg_3^{t}=0$ for all $0\leq t\leq T$, then 
\beq
   en_2^t=0, \ \ EN_2^t=0, && 0\leq t\leq T+2,\\
   hh_2^t=0, && 1\leq t\leq T+2,\\
   HH_2^t=0,  && 1\leq t\leq T+3, \\
   PTC_3^t=1, && 1\leq t\leq T+3,\ \mbox{ and }\\
   CIA_3^t=0, \ \ CIR_3^t=1, && 2\leq t\leq T+3.
\eeq
\item[(b)]
Furthermore, if $ci_3^0=0$, then also $CIA_3^1=0$ and part (a) holds
for any $T\geq0$.
\eit 
\efc

With the help of this Fact, we establish that $wg_3$ may become expressed only at the first iterate or else it
is never expressed. Thus the two timescale model provides a strong natural restriction on the formation of an
inviable state: if $wg_3^{1}=0$, then $wg_3^{T}=0$ for all $T\geq0$, implying that such trajectories will 
never converge to the broad striped pattern.

\bp{pr-wg3-T=1only}
Assume that the initial condition satisfies $wg_3^0=0$, $ptc_3^0=1$, $hh_{2,4}^0=0$, 
and $ci_3^0=1$. Then $wg_3^{T_1}=1$ and $wg_3^{T}=0$ for all $0\leq T<T_1$, only if $T_1=1$.
\ep

\bpr
To obtain $wg_3^{T}=1$ with $wg_3^{T-1}=0$ it is necessary that
\beq
   wg_3^T=CIA_3^T\mbox{ and not } CIR_3^T \ \ \Rightarrow\ \   
   CIA_3^{T}=1\ \mbox{ and } \ CIR_3^{T}=0.
\eeq
But, if $wg_3^{T}=0$ for $T=1$, then, by Fact~\ref{fc-wg3=0}, the activator $CIA_3$ is zero for $T=2,3,4$.
Then (by induction on $T$) expression of $wg_3$ is prevented at any later time.
\epr

In addition, it is possible to completely characterize the updating permutation ($\phi^1$) that leads  
to $wg_3^1=1$ and, as a consequence, exactly compute the probability of divergence (hence convergence) to
the wild type steady state (Section~\ref{sec-probability}).

\bp{pr-wg3-T=1}
Assume that assumptionts A1 and A2 hold. Assume that the initial condition satisfies 
$wg_3^0=0$, $ptc_3^0=1$ and $hh_{2,4}^0=0$.
\bit
\item[(a)] If $ci_3^0=0$, then $wg_3^{1}=0$.
\item[(b)] If $ci_3^0=1$, then $wg_3^{1}=1$ if and only if the permutation 
$\phi^1$ satisfies the following sequence among the proteins $CI$, $CIA$, $CIR$ and $PTC$:  
\beqn{eq-perm-wg3=1}
\begin{array}{cccccc}
    CIR_3 & CI_3 & & CIA_3  & & PTC_3,        \\
    & & & & & \\ 
          & CI_3 &   CIR_3  & CIA_3 & & PTC_3,  \\
    & & & & & \\ 
          & CI_3 & & CIA_3  & CIR_3 & PTC_3,
\end{array}
\eeqn
while the other proteins may appear in any of the remaining slots.
\eit
\ep

\bpr
Part (a) follows immediately from Fact~\ref{fc-wg3=0}(b).
To prove part (b), we start by noticing that, because $SLP_{3}=1$ and $wg_3^0=0$, 
\beq
   wg_3^1= CIA_3^1 \mbox{ and not } CIR_3^1,
\eeq
so that
\beq
  wg_3^{1}=1\ \ \ \Leftrightarrow\ \ \  CIA_3^{1}=1 \mbox{ and } CIR_3^{1}=0.
\eeq
Following assumptions A1-A2, the model's equations for $CIA_3^1$ and $CIR_3^1$ are given by:
\beq
  CIA_3^1 &=& CI_3^{t_a} \mbox{ and }
        [\mbox{not }PTC_3^{t_b} 
         \mbox{ or } HH_{2}^{t_c}\mbox{ or }HH_{4}^{t_d}\mbox{ or } hh_{2}^0\mbox{ or } hh_{4}^0 ], \\
  CIR_3^1 &=& CI_3^{s_a} \mbox{ and not } 
        [\mbox{not }PTC_3^{s_b}
         \mbox{ or } HH_{2}^{s_c}\mbox{ or }HH_{4}^{s_d}\mbox{ or } hh_{2}^0\mbox{ or } hh_{4}^0 ],
\eeq
where $t_a,\ldots,s_a\in\{0,1\}$ and depend on the permutation $\phi^1$.
These expressions may be simplified by observing that: 
(a) $hh_{2,4}^0=0$, and thus also
(b) $HH_{2,4}^{0,1}=0$.
Therefore,
\beq
  CIA_3^1 &=& CI_3^{t_a} \mbox{ and not } PTC_3^{t_b}, \\
  CIR_3^1 &=& CI_3^{s_a} \mbox{ and } PTC_3^{s_b},
\eeq
The values for $CI_3^{0,1}$ and $PTC_3^{0,1}$ are determined by:
\bit
\item[1.] $CI_3^0=0$ and  $CI_3^1=ci_3^0=1$,
\item[2.] $PTC_3^0=0$ and $PTC_3^1=ptc_3^0 \mbox{ or }[\cdots]=1$, since $ptc_3^0=1$,
\eit
and recall that both $CIA_i^0=0$ and $CIR_i^0=0$.
Therefore, {\it it is necessary that $CI_3$ is updated before $CIA_3$ and  
$PTC_3$ is updated after $CIA_3$ }, because otherwise $CIA_3^1=0$. 
Finally, {\it $CIR_3$ must be updated before $PTC_3$}, because otherwise
$CIR_3^1=1$. In other words: 
\beq
   t_a=1, \ \ t_b=0,\ \ s_a\in\{0,1\},\ \ s_b=0.
\eeq 
It is easy to see that any of the sequences\rf{eq-perm-wg3=1} is also sufficient to
obtain $wg_3^{1}=1$.
\epr

Finally, we will show that whenever $wg_3$ becomes expressed at time $T=1$, it is afterwards periodically 
expressed, every third step.  Such trajectories cannot converge to the wild type pattern.
In other words, initial permutations of the form\rf{eq-perm-wg3=1} are not included in the basin of attraction 
of the wild type pattern.

\bp{pr-mutant-conv}
Assume $wg_4^0=1$, $ci_1^0=0$ and $ptc_1^0=0$. For any $T\geq1$, if $wg_3^T=1$, then $wg_3^{T+3}=1$.
\ep

\bpr
Recall that, by Proposition~\ref{pr-two-stst}, $wg_4^t=1$ for all $t\geq0$. 
By Corollary~\ref{cor-var2}, $CIR_2^t=0$ for all $t\geq3$.

Now, pick any $T\geq1$ and assume that $wg_3^T=1$. Then
\beq
  WG_3^{T+1}=1\ \Rightarrow\ en_2^{T+1}=1 \ \Rightarrow\ EN_2^{T+2}=1
  \ \Rightarrow\ hh_2^{T+2}=wg_3^T=1,
\eeq
where the last implication follows from Fact~\ref{fc-hh2}.
Then (using either\rf{eq-ci34-T=0} or\rf{eq-ci34-T=1}) 
\beq
  CIA_3^{T+3} &=& \mbox{not }PTC_3^{t_a}\mbox{ or } HH_{2}^{t_b}\mbox{ or } hh_{2}^{T+2},\\
  CIR_3^{T+3} &=& PTC_3^{s_a}\mbox{ and not } HH_{2}^{s_b}\mbox{ and not } hh_{2}^{T+2},
\eeq
(where $t_a,t_b,s_a,s_b\in\{ T+2,T+3\}$, and depend on the permutation $\phi^{T+3}$).
So $hh_2^{T+2}=1$ implies
\beq
   CIA_3^{T+3}=1\ \ \mbox{ and }\ \     CIR_3^{T+3}=0,
\eeq
and therefore
$
   wg_3^{T+3}=1,
$
as we wanted to show.
\epr

Whenever $wg_3$ is not expressed, some other nodes also stabilize, after the appropriate number 
of iterations. These are summarized next.

\bc{cor-var3}
Assume that $wg_3^t=0$ for all $t\geq0$. Then $WG_3^t=0$, $en_2^t=0$ for all $t\geq1$.   
$EN_2^t=0$, $hh_2^t=0$ and $ci_2=1$ for all $t\geq2$. Finally, $HH_2^t=0$ and $CI_2=1$ for all $t\geq3$.
\ec

\subsection{Probability of convergence to wild type}
\label{sec-probability}

The wild type pattern is in fact an attractor for the asynchronous model: every trajectory which is not of 
the form\rf{eq-perm-wg3=1} converges to the wild type pattern (see Appendix~\ref{sec-attractor}).
The probability that this happens is therefore determined 
by counting all the possible  states of the form\rf{eq-perm-wg3=1}:
\beq
  \mbox{Prob(wild type) } =1 - \frac{\mbox{\# permutations as in\rf{eq-perm-wg3=1}}}
                                {\mbox{Total \# permutations}} 
\eeq
Let $L$ be the number of protein nodes to be updated at each iterate (there are 9 proteins in each of 
the four cells so $L=36$). Out of the $L$ proteins, only 4 need to satisfy one of particular 
sequences\rf{eq-perm-wg3=1} in their relative positions. So let $M_L$ be the number of possible permutations 
satisfying any of the sequences\rf{eq-perm-wg3=1}. Then
\beq
   \mbox{Prob(wild type) } = 1 - \frac{M_L\ (L-4)!}{L!}.
\eeq
The next Proposition is proved in the Appendix and shows that, in fact, this is a constant number, 
independent of $L$. That is, 
\beq
   \mbox{Prob(wild type) }=0.875.
\eeq

\bp{pr-prob}
For any $L\geq4$,
\beq
   M_{\tL}= \frac{3!}{2}\ \sum_{P=4}^{L}\;\left( 
                                   \begin{array}{c}
                                      P-1 \\ 3
                                   \end{array}  
                               \right)
          =\frac{1}{2}\ \sum_{j=1}^{L-3}\; j(j+1)(j+2)    
\eeq
and
\beq
   \frac{M_{\tL}\ (L-4)!}{L!} = \frac{1}{8}.
\eeq
\ep

\section{A Markov chain process}
\label{sec-markov}

\begin{figure}[htbp]
\centerline{\scalebox{0.7}[0.7]{\includegraphics{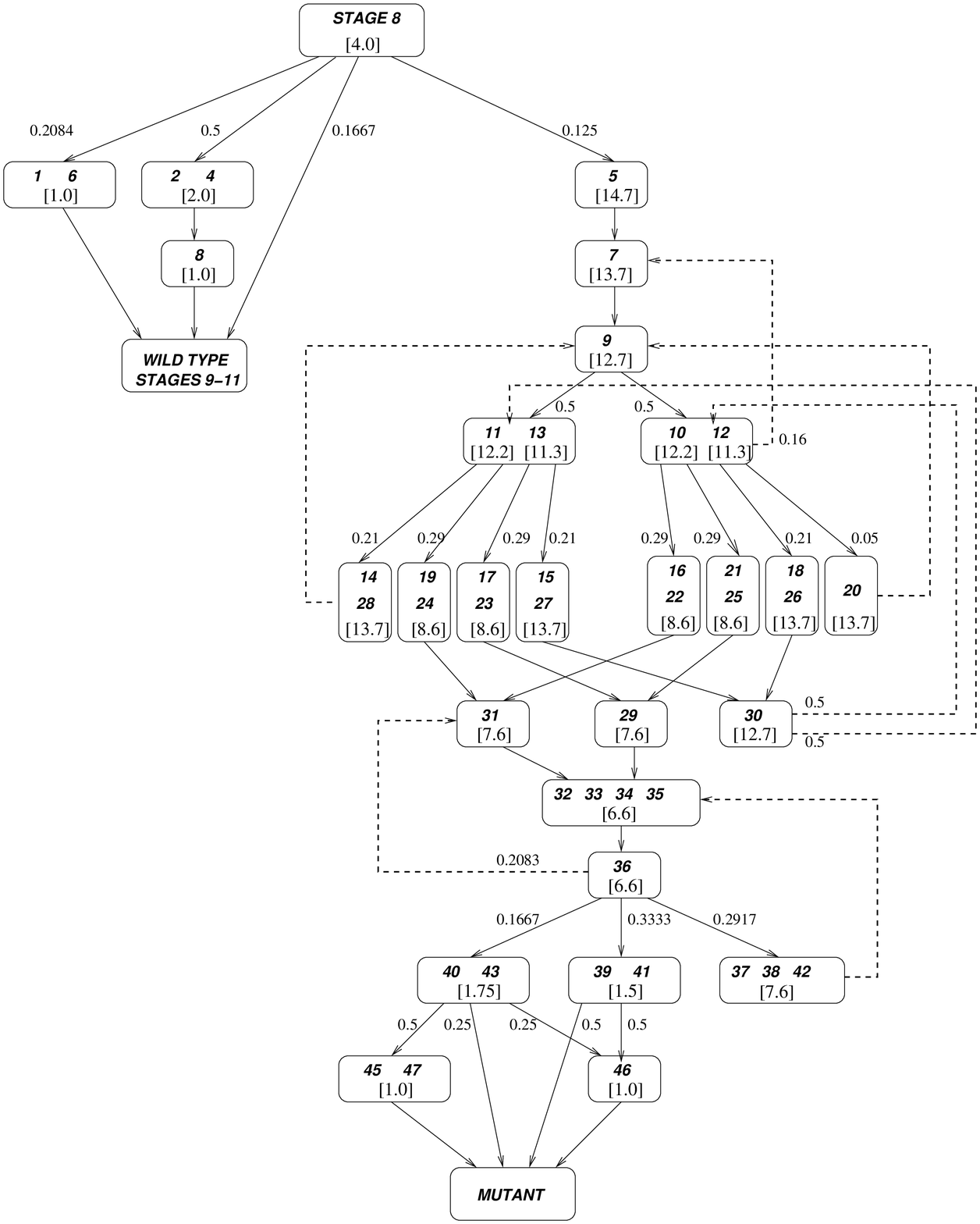} }}
\caption{Robustness of the regulatory network modeled with the two-timescale algorithm.
There are 48 states reachable from the wild type initial state.
The arrows are labeled by the transition probabilities between states
(if unlabeled, the  probabilities are 1), and the expected times to absorption
into the  corresponding steady state are indicated between square brackets.}
\label{fig-diag}
\end{figure}

As a Boolean model, there are only a finite set, say $\S$, of distinct states
(in the total state space $\{0,1\}^N$) reachable by the system. 
Starting from any state $\S_a\in\S$, each permutation $\phi$ of 
$\{1,\ldots,N\}$ takes the system to some other state $\S_b\in\S$.
It is possible to theoretically identify all the distinct intermediate
and final states of the system as well as all the possible transitions
after one iteration. 
Thus the asynchronous algorithm consisting of the $N$ node 
functions\rf{eq-inode} together with assumptions (A1), (A2) and (A3)
may be characterized as a Markov chain process, by identifying the $d$
distinct states
\beq
  \S=\{\S_1,\S_2,\ldots,\S_d\},
\eeq
and the $d\times d$ transition matrix $P$, where 
\beq
  P_{ij}=\mbox{ probability of a transition from state $\S_i$
    to state $\S_j$}.
\eeq
The probabilities $P_{ij}$ are simply the fraction of the total
number of  permutations that (in one iterate) transform the state 
$\S_i$ into state $\S_j$. The matrix $P$ is a stochastic matrix, since
all its rows add up to 1.
A state $\S_a$ with the property that all permutations leave the state
unchanged (that is, $P_{aa}=1$ and $P_{aj}=0$ for $j\neq0$) is
called an {\em absorption state} of the Markov chain, and it is also 
a steady-state of system\rf{eq-inode}.
In the asynchronous model there are only two absorption states,
corresponding to the wild type and broad {\em wingless} stripe mutant patterns, 
as described above.
Isolating the two rows and columns that correspond to these
absorption states, the transition matrix may be partitioned as
\beq
   P=\left[\begin{array}{cc}
             P_a & 0 \\
             R_a & \bar P 
           \end{array} \right],
\eeq
where $\bar P$ is of size $(d-2)\times(d-2)$. It is a well know 
result(see any standard book on probability theory, for instance~\cite{feller}) 
that $I-\bar P$ is an invertible matrix, and 
\beq
   \left[\begin{array}{c}
          \bar T_1 \\ \bar T_2 \\ \vdots \\ \bar T_{d-2}
         \end{array}\right]
   =(I-\bar P)^{-1} \left[\begin{array}{c}
                         1 \\ 1 \\ \vdots \\ 1
                     \end{array}\right],
\eeq
or
$
   \bar T_i = 1 + \sum_{j=1}^{d-2}\ p_{ij}\;\bar T_j. 
$
The values $\bar T_i$ provide an estimated time for absorption when the chain starts from state $\S_i$ 
(if $a$ is an absorption state, then $\bar T_a=0$).
In Figure~\ref{fig-diag}, a schematic diagram of transitions is shown, together with probabilities and 
estimated times for absorption. This diagram was obtained from a simulation starting with the initial 
wild type pattern (observed at stage 8 of the embrionic development), and following the assumptions (A1)-(A3), 
as well as the additional
\bit
\item[(A4)] Each protein or gene is updated simultaneously in the four cells,
\eit
meaning that there is no cell-to-cell variation in the duration of molecular processes.
In this case, the total number of possible permutations is a manageable $7!\times5!=604800$:
\beq
   \Phi=\{\phi=(\phi_{\tProt},\phi_{\tmRNA}): & & 
         \phi_{\tProt} \mbox{ is a permutation of $\{1,2,3,4,5,6,7\}$ },\\ 
        && \phi_{\tmRNA} \mbox{ is a permutation of $\{1,2,3,4,5\}$ }\}
\eeq
The total number of distinct states of the Markov chain (under assumptions (A1)-(A4)) 
is $d=48$. The transition probabilities matrix was computed exactly, by counting 
all the $7!\times5!$ transitions from each of the 48 states.

It is clear from this diagram that the decision between convergence to the wild type or the mutant patterns 
is indeed decided at the first iteration, in agreement with Propositions~\ref{pr-two-stst} and~\ref{pr-wg3-T=1only}.
Furthermore, the diagram shows the possibility of periodic oscillations (of period at least three) in the 
mutant branch (see also Proposition~\ref{pr-mutant-conv}). 
Although in practice the probability of a limit cycle is very small, this prevents (theoretical) convergence 
to the mutant state, and considerably increases the absorption times to the mutant state. 
The robustness of the two timescale model is illustrated by the fast convergence to the wild type pattern 
(expected time to convergence is 4 steps), contrasted with the long and oscillation-strewn path toward the 
broad striped pattern (expected time to convergence is 15 steps).

\section{Identifying minimal pre-patterns}

A necessary condition for convergence to the wild type is that $ptc_3^0=1$.
Otherwise the trajectory immediately fails to enter the basin of attraction of the
will type state:

\bfc{fc-wg3=1} 
Assume $ptc_3^0=0$. Then $wg_3^T=1$, for some $T\in\{1,2,3\}$.
\efc

\bpr
Note that $ptc_3^0=0$ implies $PTC_3^1=0$.
Using\rf{eq-en-hh34} and\rf{eq-ci34-T=0},\rf{eq-ci34-T=1}, 
the equations for $ptc_3$ and $CIA_3$ simplify to
\beq
   CIA_3^t &=& \mbox{not }CIR_3^t=
      \mbox{not }PTC_3^{t_a}\mbox{ or }HH_2^{t_b}\mbox{ or }hh_2^{t-1}, \ \ \ t\geq\tau \\
   ptc_3^t &=& CIA_3^t\mbox{ and not }CIR_3^t=CIA_3^t,\ \ \ t\geq1 \\
   PTC_3^t &=& ptc_3^{t-1} \mbox{ or }
      [PTC_3^{t-1}\mbox{ and not }HH_2^{t_a}\mbox{ and not }hh_2^{t-1},\ \ \ t\geq1,
\eeq
where $\tau=2$ (respectively, $\tau=3$), if $ci_3^0=1$ (respectively, $ci_3^0=0$). 
Consider first the case $ci_3^0=1$. 
The activator protein will be turned on either at the first or second iterations:
$CIA_3$ may become activated at $t=1$ because $PTC_3^1=0$, (if the permutation $\phi^1$
is such that $CI_3$ is updated before $CIA_3$). If $CIA_3$ is not actived at $t=1$, then
it certainly is actived at $t=2$ (because $CIA_3^1=0$ implies $ptc_3^1=0$ and $PTC_3^2=0$).
A similar argument shows that $CIR_3^{1,2}=0$.

Consider next the case $ci_3^0=0$: we have $CIA_3^1=CIR_3^1=0$, and $CIA_3$ is turned on at the 
second or third iterations, by a similar argument as before.
Therefore, the {\em wingless} gene is also expressed after $CIA_3$, at $T=1,2$ or 3.
\epr

Another necessary condition for convergence to wild type, is that 
\beq
   wg_4^0=1\ \mbox{ or }\ en_1^0=1\ \mbox{ or }\ ci_4^0=1.
\eeq
Otherwise, the trajectories cannot converge to the wild type nor to the mutant steady states,
In this case, the only possible steady state is a ``lethal'' state, where expression of 
$PTC$, $ci$, $CIA$ and $CIR$ is ubiqitous and all others are absent.

\bp{fc-PTC1}
Assume $wg_{3,4}^0=0$, $en_1^0=0$, $ptc_3^0=1$ and $ci_{3,4}^0=0$. Then 
\bit
\item[(a)] $wg_3=0$, for all $t\geq0$.
\item[(b)] $PTC_1^t=1$ for all $t\geq4$.
\item[(c)] $wg_4=0$ for infinitely many $t$. 
\eit
\ep

\bpr
Part (a) follows from Fact~\ref{fc-wg3=0}(b), for any $T\geq0$:
\beq
   wg_3^t=0,\ \ t\leq T \ \ \Rightarrow\ \ CIA_3=0,\ \ 0<t\leq T+3.
\eeq
Thus, by induction, $wg_3^t=0$ for $t\geq0$.

Part (a) and Fact~\ref{fc-hh2} imply $hh_2^t=0$ and $HH_2^t=0$ for all $t>0$, so that 
\beq
   ptc_1^t &=& CIA_1^t\mbox{ and not }CIR_1^t\mbox{ and not }EN_1^t \\
   PTC_1^t &=& ptc_1^{t-1}\mbox{ or }PTC_1^{t-1},\ \ t\geq1 \\
   CIA_1^t &=& CI_1^{t_a}\mbox{ and not }PTC_1^{t_b},\ \ t\geq1.
\eeq
If $ptc_1^0=1$, then it is clear that $PTC_1^t=1$ for all $t\geq1$. 
Consider the case $ptc_1^0=0$. Then $PTC_1=0$ as long as $ptc_1=0$.  
Note that $ci_4^0=0$ implies $CI_4^1=0$, $CIA_4^1=0$ and also $wg_4^1=0$. 
Then $wg_4^{0,1}=0$ and $en_1^{0,1}=0$ imply $EN_1^{1,2,3}=0$ and $CI_1^{2,3,4}=1$.
So, either at $T=2$ or $T=3$, we will have $CIA_1^T=1$, and therefore also 
$ptc_1^T=1$ and $PTC_1^{T+1}=1$.
Thus, $PTC_4^t=1$ for $t\geq4$, proving part (b).

Finally, to argure by contradiction, suppose that $wg_4^t=1$ for $t\geq T_a$. Then
by Corollary~\ref{cor-var2}, $PTC_1^t=1$ for $t\geq T_b> T_a$, which contradicts part (b).
Hence, $wg_4$ cannot become permanently on.
\epr

By Proposition~\ref{pr-two-stst}, together with Fact~\ref{fc-wg3=0}(b), a 
{\em sufficient} condition for convergence to wild type is
\beq
   wg_4^0=1,\ \ ptc_3^0=1,\ \ ptc_1^0=0,\ \ ci_{1,3}^0=0. 
\eeq
Another {\em sufficient} condition (which allows the presence 
of {\em cubitus} in the third cell) is 
\beq
    wg_4^0=1,\ \ ptc_3^0=1,\ \ PTC_3^0=1.
\eeq
The argument in the proof of Proposition~\ref{pr-wg3-T=1}, 
shows that, if $PTC_3^0=1$ then $wg_3^1=0$. Then, by 
Proposition~\ref{pr-wg3-T=1only}, it follows that $wg_3^t=0$
for all times.

In conclusion, while the wild type initial state allows for an ambiguity in the final states, we find that a remarkably 
minimal prepattern, consisting of $wg_4$ and $ptc_3$, is sufficient to guarantee the convergence to the wild 
type steady state. In other words, the initiation of two genes in two cells is enough to compensate for 
initiation delays in any and all other genes, irrespectively of the variations in individual synthesis and 
decay processes. This suggests a remarkable error correcting ability of the segment polarity gene control 
network.

\section{Conclusions}
In summary, we proposed  an intuitive and practical way of introducing stochasticity in
qualitative models of gene regulation. We explored three possible ways of incorporating
the variability of transcription, translation, post-translational modification and decay
processes (see Table \ref{table_comp_alg} for a comparison between
the synchronous and three asynchronous algorithms). Applying our methods on a previously introduced
model of the {\it Drosophila} segment polarity genes gave us new insights into
the dynamics and function of the interactions among the segment polarity genes
and, through it, into the robustness of the embryonic segmentation process.
Our results suggest that unrestricted variability in synthesis/decay/transformation 
timescales can lead to a divergence from the wild 
type development process, with an expected divergence probability of $45\%$. On the other hand,
if the duration of post-translational transformations is consistently less than the duration
of transcription, translation and mRNA/protein half-lives, the wild type steady
state will be achieved with a high probability, despite significant variability
in individual process durations. We find that a remarkably sparse
prepattern is sufficient to ensure the convergence to the wild type steady state of these genes. 
This dual behavior, robustness to changes in the initial state
but fragility with respect to temporal variability, is reminiscent of Highly Optimized Tolerance,
a feature of highly structured, non-generic complex systems with robust, yet fragile external
dynamics~\cite{cd02}. Similar robust-yet-fragile features have also been found in the 
of structure of diverse networks~\cite{jtaob00,jmbo01,ajb00}.

\begin{table}[htb]
\caption{Comparison of synchronous and asynchronous algorithms.}
\label{table_comp_alg}
\centerline{
\begin{tabular}{lllll}
\hline
& Synchronous & Totally Asynchronous & Random Order & Two-timescale\\
\hline\hline
Assume & Nodes are updated  & The time between &
Each node is  & In each time \\
  &  at multiples  &  updates is   &
 updated at a randomly   &  interval proteins are\\
 & of the unit  & perturbed in a  & selected point of the &  updated first, \\
& time interval. & range $\pm \epsilon$. & unit time 
interval. & then mRNAs.\\
\hline
Update & $T^k = k$  & $T^k = T^{k-1}+1+\epsilon r^k$ & $T^k = k-1+\frac 1 N \phi^k$ 
& $T^k = k-1+\frac 1 N \phi^k_{tt} $\\
&  &  & $\phi^k$ - node permutation & $\phi^k_{tt} \in (\phi_{\tProt}^k,\phi_{\tmRNA}^k)$\\ 
\hline
Pros & Correctly identifies & Allows for unlimited & Does not depend & Allows separation 
 \\
 &  all steady states. & variability in & on any perturbation  & of post-translational\\
 & Can be solved & process durations. & parameter $\epsilon$. &  and pre-translational
 \\
 & analytically. & & &  processes.\\
\hline
Cons & Dynamics is & \multicolumn{2}{l}{Can have unrealistically short transcription} & 
Only useful when \\
& unrealistic. & \multicolumn{2}{l}{and translation times.} & process durations \\
& & & & can be separated.\\
\hline
Results & Prepattern errors  & \multicolumn{2}{l}{Divergence from the wild type process is 
possible.} & Development is stable if \\
& can be corrected. & \multicolumn{2}{l}{Cause: imbalance between two 
transcription factors.} & PTC 
is prepatterned.\\
\hline
\end{tabular}}
\end{table}

All our algorithms concur in suggesting that
the divergence from wild type can be attributed to an imbalance between the
two opposing Cubitus Interruptus transcription factors (CIA, CIR) in the posterior half of the 
parasegment. Thus the complementary regulation and pattern of these opposing transcription
factors (Aza-Blanc and Kornberg 1999) is a vital requirement for the correct functioning
of the segment polarity gene network.
The totally asynchronous algorithm predicts that  perturbations to the 
post-translational modification of Cubitus Interruptus can have effects as severe 
as mutations: a transient overexpression of CIR leads to the pattern 
with no segmentation, while transient expression of CIA and not CIR leads to the broad striped pattern.
With the two timescale algorithm we find that the condition for the divergence from the wild type
pattern is that, in the third cell of the parasegment, the post-translational modification
of CI precedes the synthesis of the Patched protein. The biological realization
of this condition appears unlikely, since PTC is documented as being ubiquitously expressed during cellularization 
(stage 5)~\cite{tnmi93}, while the post-translational modification
of CI requires SMO that is only weakly expressed until stage 8~\cite{azn00}. Our model
predicts that if for any reason the PTC protein is absent in the period when
the pair-rule proteins decay and the regulation between the segment polarity genes
starts, the wild type expression pattern is unreachable."

Our methods combine the benefits of discrete-state models with a continuum in timescales. In 
the absence of quantitative information, we considered every possible timescale or update order, 
but as the two-timescale model demonstrates, existing information can be easily incorporated.
We were able to describe the system in a rigorous mathematical way, to 
identify the relatively few types of behavior possible in the system (the attractors in 
state space) and to
theoretically prove the convergence toward these states. Our results underscore that 
predictive mathematical modeling is possible despite the scarcity of quantitative information on gene 
regulatory processes.

\section*{Acknowledgements} 
The work of M.C. was supported in part by NIH Grants P20 GM64375 and Aventis. 
R.A. gratefully acknowledges an Alfred P. Sloan Research Fellowship. 
The work of E.D.S. was supported in part by NSF Grant CCR-0206789 and
NIH Grants P20 GM64375 and R01 GM46383.

\appendix
\section{Additional Proofs}

{\em Proof of Fact~\ref{fc-CIR1}:}
For $T=0$, the statement follows directly from the model's equations and by using the
assumptions A1-A2 repeatedly:
\beq
  WG_4^{1}=1,\ \ en_1^1=1,\ \ EN_1^{2}=1, 
\eeq
as well as 
\beq
  ci_1^{0,2}=0,\ \ CI_1^{0,1,3}=0,\ \ CIA_1^{0,1}=0,\ \ CIR_1^{0,1}=0,\ \ PTC_1^{0,1}=0.
\eeq
We have (using\rf{eq-en-hh34}) 
\beq
  ptc_1^{t} &=& CIA_1^{t}\mbox{ and not }CIR_1^{t}\mbox{ and not }EN_1^{t} \\
  PTC_1^{t} &=& ptc_1^{t-1}\mbox{ or }[PTC_1^{t_b}\mbox{ and not }HH_2^{t_c}],\\
  CIR_1^{t} &=& CI_1^{t_a}\mbox{ and }PTC_1^{t_b}\mbox{ and not }HH_2^{t_c}\mbox{ and not }hh_2^{t-1}, 
\eeq
so that we conclude
\beq
  ptc_1^{0,1,2}=0,\ \ PTC_1^{2,3}=0,\ \ CIR_1^{2,3}=0.
\eeq
We next prove the Fact by induction. First note that, for any $t\geq0$:
\beq
   wg_4^{t}=1 \ \Rightarrow\ WG_4^{t+1}=1 \ \Rightarrow\ en_1^{t+1}=1
   \ \Rightarrow\ EN_1^{t+2}=1
\eeq
and this implies 
\beqn{eq-ciCI1}
   ci_1^{t+2}=0 \ \Rightarrow\ CI_1^{t+3}=0. 
\eeqn
Now assume that the Fact holds for some $T\geq1$ and that 
\beq
     wg_4^{t}=1,\ \   0\leq t\leq T.
\eeq
By the induction hypothesis, we know that
\beq
     CI_1^{t}=0, \ 3\leq t\leq T+2,\ \ \mbox{ and }\ \ CIR_1^{t}=0, \ 0\leq t\leq T+2,
\eeq
By\rf{eq-ciCI1}, $wg_4^T=1$ implies $CI_1^{T+3}=0$, and this together
with $CI_1^{T+2}=0$ also guarantees that $CIR_1^{T+3}=0$, as we wanted to show.
\qed

{\em Proof of Fact~\ref{fc-wg3=0}:}
To prove part (a), assume that $wg_3^{t}=0$ for all $0\leq t<T$, with $T\geq1$.
Since $en_2^0=0$, then $EN_2^{0,1}=0$ and $hh_2^1=0$, $HH_2^{1,2}=0$. 
For $t\geq3$ apply Fact~\ref{fc-hh2} to obtain
the desired value for $hh_2$ and $HH_2$.
Note that $ptc_3^0=1$ implies $PTC_3^1=1$ and, together with
$HH_2^{1,2}=0$, also $PTC_3^2=1$; then the value of $PTC_3$ 
follows from Fact~\ref{fc-PTC3}.
For $T\geq2$, and using\rf{eq-en-hh34} and\rf{eq-ci34-T=1}), we have 
\beq
  CIA_3^t &=&\mbox{not }PTC_3^{t_b}\mbox{ or } HH_{2}^{t_c}\mbox{ or } hh_{2}^{t-1} \\
  CIR_3^t &=& PTC_3^{s_b}\mbox{ and not } HH_{2}^{s_c}\mbox{ and not } hh_{2}^{t-1}
\eeq
so, the values for $hh_2$, $HH_2$ and $PTC_3$, indeed imply that
$CIA_3^t=0$ and $CIR_3^t=1$, for $2\leq t\leq T+3$. 

To prove part (b), note that if $ci_3^0=0$, then also $ci_3^{0,1}=0$ and hence $CIA_3^1=0$.
But now $CIA_3^1=0$ together with $wg_3^0=0$ immediately imply that
$wg_3^1=0$, and therefore, the results in part (a) are valid for all $T\geq0$.
\qed

\bfc{fc-hh2}
\beq
  hh_2^{T+2}=HH_2^{T+3}=wg_3^T\mbox{ and not} CIR_2^{T+2},\
  \mbox{ for all }\ T\geq0.
\eeq
In particular, if $CIR_2^t=0$ for all $t\geq3$, then
$hh_2^{T+2}=HH_2^{T+3}=wg_3^T$, for all $T\geq1$.
\efc

\bpr
Given any $t\geq0$ it is easy to see that
\beq
   WG_3^{t+1} &=& wg_3^t   \\
   en_2^{t+1} &=& WG_1^{t+1} \mbox{ or } WG_3^{t+1} \equiv WG_3^{t+1}, \\
   EN_2^{t+2} &=& en_2^{t+1}, \\
   hh_2^{t+2} &=& EN_2^{t+2} \mbox{ and not } CIR_2^{t+2}, \\
   HH_2^{t+3} &=& hh_2^{t+2},
\eeq 
where the equation for $en_2^{t+1}$ follows from\rf{eq-wg12}.
\epr

\bfc{fc-PTC3}
$PTC_3^T=1$ and $PTC_3^{T+1}=0$, for some $T>0$, only if $wg_3^{t}=1$ for some 
$t\in\{T-3,T-2\}$ (chosen according to the permutation $\phi^{T-2})$.

\efc

\bpr
To see this, simply notice that
\beq
   PTC_3^{T+1} &=& ptc_3^T \mbox{ or }[PTC_3^{T} \mbox{ and not } HH_2^{t_a} \mbox{ and not } HH_4^{t_b}] \\
               &=& ptc_3^T \mbox{ or }[PTC_3^{T} \mbox{ and not }
     HH_2^{t_a}]\\
               &=& ptc_3^T \mbox{ or }[PTC_3^{T} (\mbox{ and not }
wg_3^{t_a-3}\mbox{ or } CIR_2^{t_a-1}] 
\eeq
because from\rf{eq-en-hh34} $HH_4^{T}=0$, and by Fact~\ref{fc-hh2}. 
Note that $t_a\in\{T,T+1\}$, depending on the permutation $\phi^{T+1}$.
So, for $PTC_3$ to vanish it is necessary that both $ptc_3^T=0$ and
$wg_3^{t_a-3}=1$.
\epr

\section{Attractiveness of the wild type pattern}
\label{sec-attractor}

Assuming that the trajectory is not of the form\rf{eq-perm-wg3=1}, the only accessible steady state is the 
wild type. In this case, to establish convergence of the trajectory, it is enough to show that each node 
attains a constant value after a finite number of iterates.
And in fact, from Propositions~\ref{pr-two-stst},~\ref{pr-wg3-T=1only} and 
Corollaries~\ref{cor-var},~\ref{cor-var2}, all the nodes become fixed  after at most $t$ iterates, as indicated:
\beq
   wg_{1,2}=0,\ \ WG_{1,2}=0,\ \ t\geq0,  & & \mbox{~\rf{eq-wg12}} \\
   wg_4=1,\ \ WG_4=1,\ \ t\geq1,          & & \mbox{ Proposition~\ref{pr-two-stst}}\\
   wg_3=0,\ \ WG_3=0,\ \ t\geq1,          & & \mbox{ Proposition~\ref{pr-wg3-T=1only}}\\
   en_1=1,\ \ EN_1=1,\ \ t\geq2,          & & \mbox{ Corollary~\ref{cor-var}} \\
   en_2=0,\ \ EN_2=0,\ \ t\geq2,          & & \mbox{ Corollary~\ref{cor-var3}} \\
   en_{3,4}=0,\ \ EN_{3,4}=0,\ \ t\geq0,  & & \mbox{~\rf{eq-en-hh34}} \\
   hh_1=1,\ \ HH_1=1,\ \ t\geq3,          & & \mbox{ Corollary~\ref{cor-var}} \\
   hh_2=0,\ \ HH_2=0,\ \ t\geq3,          & & \mbox{ Corollary~\ref{cor-var3}} \\
   hh_{3,4}=0,\ \ HH_{3,4}=0,\ \ t\geq0,  & & \mbox{~\rf{eq-en-hh34}}
\eeq
\beq
   ci_1=0,\ \ CI_1=0,\ \ t\geq3,          & & \mbox{ Corollary~\ref{cor-var}} \\
   ci_2=1,\ \ CI_2=1,\ \ t\geq3,          & & \mbox{ Corollary~\ref{cor-var3}} \\
   ci_{3,4}=1,\ \ CI_{3,4}=1,\ \ t\geq2,  & & \mbox{~\rf{eq-ci34-T=0},~\rf{eq-ci34-T=1}}  
\eeq
\beq
   ptc_1=0,\ \ PTC_1=0,\ \ t\geq0,        & & \mbox{ Corollary~\ref{cor-var}} \\
   CIA_1=0,\ \ CIR_1=0,\ \ t\geq4,        & & \mbox{ Corollary~\ref{cor-var}} \\
   CIA_3=0,\ \ CIR_3=1,\ \ t\geq2,        & & \mbox{ Fact~\ref{fc-wg3=0}} \\
   ptc_3=0,\ \ PTC_3=1,\ t\geq1,     & & \mbox{ Fact~\ref{fc-wg3=0}} \\
   CIA_2=1,\ \ CIR_2=0,\ \ t\geq4,        & & \ CI_2=1 \\
   ptc_2=1,\ \ PTC_2=1,\ \ t\geq5,        & & \ CIA_2=1,\ \ CIR_2=0,\ \ EN_2=0 \\
   CIA_4=1,\ \ CIR_4=0,\ \ t\geq3,        & & \ wg_4=1 \\
   ptc_4=1,\ \ PTC_4=1,\ \ t\geq5,        & & \ CIA_4=1,\ \ CIR_4=0.       
\eeq
This is indeed a complete characterization of the wild type steady state.

{\it Proof of Proposition~\ref{pr-prob}:}
Let $P$, $A$, $C$ and $R$ ($\leq L$) denote the positions of $PTC_3$, $CIA_3$,
$CI_3$ and $CIR_3$, respectively. Then, from\rf{eq-perm-wg3=1} it is
easy to see that
\beq
   P\in\{4,5,6,\ldots,L\},\ \ 
   A\in\{2,3,4,\ldots,P-1\},\ \ 
   C\in\{1,2,3,\ldots,A-1\},\ \
   R\in\{1,\ldots,P-1\}\setminus\{P,A,C\}.
\eeq
To derive a formula for $M$, we note that, for each pair of values
$P$, $A$, the number of possible combinations of $C$ and $R$ is:
\beq
   \mbox{--- --- --- --- A --- R --- --- P}: &&  \ \ (A-1)(P-1-A)\\ 
      \\
   \mbox{--- R --- --- A  --- --- --- --- P}: && \  \  (A-1)(A-2),\\ 
\eeq
respectively for sequences of the form $CARP$ (top), or $CRAP$ and
$RCAP$ (bottom). Therefore, summing over all posible $A$ and $P$:
\beq
   M_{\tL} &=& \sum_{P=4}^{L} \sum_{A=2}^{P-1} \ [(A-1)(P-1-A) + (A-1)(A-2)] \\
    &=& \sum_{P=4}^{L} \sum_{A=2}^{P-1} \ (A-1)(P-3) \\
    &=& \frac{1}{2}\sum_{P=4}^{L}\ (P-3)(P-2)(P-1) \\
    &=&  \frac{1}{2}\sum_{j=1}^{L-3}\ j(j+1)(j+2).
\eeq
Now, for $L=4$,
\beq
   \frac{M_{\tL}\ (L-4)!}{L!}=\frac{1}{2}\;3!\ \frac{0!}{4!}=\frac{1}{2}\frac{1}{4}=\frac{1}{8}. 
\eeq
Assume now that the equality is true for $L-1$:
\beq
   \frac{M_{\mbox{\tiny $L-1$}}\ (L-5)!}{(L-1)!}=\frac{1}{8}.
\eeq
Then
\beq
   M_{\tL} &=& M_{\mbox{\tiny $L-1$}} + \frac{1}{2}(L-3)(L-2)(L-1) \\
           &=& \frac{1}{8}\frac{(L-1)!}{(L-5)!} +(\frac{1}{2}L-3)(L-2)(L-1) \\
          &=& \frac{1}{8}(L-1)(L-2)(L-3)(L-4) +(\frac{1}{2}L-3)(L-2)(L-1) \\
          &=& (L-1)(L-2)(L-3)\left[\frac{1}{8}(L-4)+\frac{1}{2}\right] \\
          &=& (L-1)(L-2)(L-3)\left[\frac{1}{8}L\right]
              =\frac{1}{8}\;\frac{L!}{(L-4)!},
\eeq  
just as we wanted to show.
\qed

\section{State aggregation in the Markov chain}
The tables below show the complete transition probabilities $p_{ij}$
(when not indicated, the transition probabilities are equal to 1).
The states numbered {\em 3} and {\em 44} denote, respectively, 
the wild type and the mutant state. The initial condition was
numbered {\em 48}.
The first table shows the complete transition probabilities at step 1, 
from the wild type initial condition.

Thus the probability shown in the diagram for the transition from
the initial state to the aggregated state \fbox{{\em 1\ \ 6}} was
obtained by adding $p_{48,1}+p_{48,6}=0.1667+0.0417=0.2084$.
A more complex aggregation formula was used for the transition
\beq
  \mbox{\fbox{{\em 10\ \ 12 }} $\ \to\ $ \fbox{{\em 16\ \ 22}}}:
\ \ \ \ 
 & &   \frac{1}{2}(p_{10,16}+p_{10,22})
      +\frac{1}{2}(p_{12,16}+p_{12,22}) \\
 &=& \frac{1}{2}(0.2083+0.0417+0.3+0.0333)=0.29165\approx0.29. 
\eeq
The normalization by $1/2$ is justified by the fact that the
transition from {\em 9} and {\em 30} to either {\em 10} or {\em 12} is the same.

\vspace{3mm}
\begin{center}
\begin{tabular}{|c|c|c|c|c|c|}
\hline
  \multicolumn{6}{|c|}
  {From initial wild type state to:} \\ 
\hline 
  {\em 1} & {\em 2} & {\em 3} & {\em 4} & {\em 5} & {\em 6} \\ 
\hline
  0.1667  & 0.357   & 0.1667  & 0.125   & 0.125   & 0.0417 \\
\hline
\end{tabular}
\end{center}

\vspace{3mm}
\begin{center}
\begin{tabular}{|c|c|c|c|}
\hline
  \multicolumn{4}{|c|}
  {From {\em 9} to:} \\ 
\hline 
  {\em 10} & {\em 11} & {\em 12} & {\em 13} \\ 
\hline
  0.25  & 0.25   & 0.25  & 0.25   \\
\hline
\end{tabular}
\end{center}

\vspace{3mm}
\begin{center}
\begin{tabular}{|c|c|c|c|c|c|c|c|c|}
\hline
  &  {\em 7} & {\em 16} & {\em 18} & {\em 20} 
  & {\em 21} & {\em 22} & {\em 25} & {\em 26}\\ 
\hline
{}From {\em 10} to:
  & 0.2083   & 0.2083   & 0.125    & 0.0417
  & 0.125    & 0.0417   & 0.125    & 0.125   \\
\hline
{}From {\em 12} to:
  & 0.1167   & 0.3      & 0.1167   & 0.05
  & 0.1333   & 0.0333   & 0.2      & 0.05   \\
\hline
\end{tabular}
\end{center}

\vspace{3mm}
\begin{center}
\begin{tabular}{|c|c|c|c|c|c|c|c|c|}
\hline
  &  {\em 14} & {\em 15} & {\em 17} & {\em 19} 
  & {\em 23} & {\em 24} & {\em 27} & {\em 28}\\ 
\hline
{}From {\em 11} to:
  & 0.2083   & 0.125    & 0.125    & 0.2083
  & 0.125    & 0.0417   & 0.125    & 0.0417   \\
\hline
{}From {\em 13} to:
  & 0.1167   & 0.1167   & 0.1333   & 0.3
  & 0.2      & 0.0333   & 0.05     & 0.05   \\
\hline
\end{tabular}
\end{center}

\vspace{3mm}
\begin{center}
\begin{tabular}{|c|c|c|c|}
\hline
  \multicolumn{4}{|c|}
  {From {\em 29} or {\em 31} to:} \\ 
\hline 
  {\em 32} & {\em 33} & {\em 34} & {\em 35} \\ 
\hline
  0.25  & 0.25   & 0.25  & 0.25   \\
\hline
\end{tabular}
\ \ \ \ 
\begin{tabular}{|c|c|c|c|}
\hline
  \multicolumn{4}{|c|}
  {From {\em 30} to:} \\ 
\hline 
  {\em 10} & {\em 11} & {\em 12} & {\em 13} \\ 
\hline
  0.25  & 0.25   & 0.25  & 0.25   \\
\hline
\end{tabular}
\end{center}

\vspace{3mm}
\begin{center}
\begin{tabular}{|c|c|c|c|c|c|c|c|}
\hline
  \multicolumn{8}{|c|}
  {From {\em 36} to:} \\ 
\hline
    {\em 31} & {\em 37} & {\em 38} & {\em 39} 
  & {\em 40} & {\em 41} & {\em 42} & {\em 43} \\ 
\hline
    0.2083   & 0.125    & 0.125    & 0.2083
  & 0.125    & 0.125    & 0.0417   & 0.0417   \\
\hline
\end{tabular}
\end{center}

\vspace{3mm}
\begin{center}
\begin{tabular}{|c|c|c|c|}
\hline
  \multicolumn{4}{|c|}
  {From {\em 37} or {\em 38} or {\em 42} to:} \\ 
\hline 
  {\em 32} & {\em 33} & {\em 34} & {\em 35} \\ 
\hline
  0.25  & 0.25   & 0.25  & 0.25   \\
\hline
\end{tabular}
\ \ \ \
\begin{tabular}{|c|c|}
\hline
  \multicolumn{2}{|c|}
  {From {\em 39} or {\em 41} to:} \\ 
\hline 
  {\em 44} & {\em 46}  \\ 
\hline
    0.5    & 0.5    \\
\hline
\end{tabular}
\ \ \ \
\begin{tabular}{|c|c|c|c|}
\hline
  \multicolumn{4}{|c|}
  {From {\em 40} or {\em 43} to:} \\ 
\hline 
  {\em 44} & {\em 45} & {\em 46} & {\em 47} \\ 
\hline
  0.25  & 0.25   & 0.25  & 0.25   \\
\hline
\end{tabular}
\end{center}


\begin{thebibliography}{100}
\bibitem[Albert \ea, 2000]{ajb00}
Albert, R., Jeong, H. \& Barab\'asi, A.-L., 2000. 
\titleref{Error and attack tolerance in complex networks.}
\pr{Nature}, \nr{406}, 378-382. 
\bibitem[Albert\- and\- Othmer, 2003]{ao03}
Albert, R. \& Othmer, H. G., 2003.
\titleref{The topology of the regulatory interactions predicts the
expression pattern of the {\it Drosophila} segment polarity genes.}
\pr{J. Theor. Biol.} \nr{223}, 1-18. 
\bibitem[Alcedo \ea, 2000]{azn00}
Alcedo, J., Zou, Y. \& Noll, M., 2000.  
\titleref{Posttranscriptional regulation of smoothened is part of a 
self-correcting mechanism in the hedgehog signaling System.} 
\pr{Molecular Cell} \nr{6}, 457-465.
\bibitem[Alon \ea, 1999]{asml99}
Alon, U., Surette, M., Barkai, N. and Leibler, S., 1999.
\titleref{Robustness in Bacterial Chemotaxis.}
\pr{Nature} \nr{397}, 168-171.
\bibitem[Aza-Blanc \& Kornberg 1999]{ak99} 
Aza-Blanc, P. \&  Kornberg, T. B. (1999)
\titleref{Ci, a complex transducer of the Hedgehog
signal}, \pr{Trends in Genetics} {\bf 15}, 458-462.
\bibitem[Bejsovec and Wieschaus, 1993]{bw93} 
Bejsovec, A. \& Wieschaus, E., 1993. 
\titleref{Segment polarity gene interactions modulate epidermal
patterning in {\it Drosophila} embryos.} \pr{Development} 
\nr{119}, 501-517.
\bibitem[Bernot \ea, 2004]{bcrg04}
Bernot, G., Comet, J.-P., Richard, A. \& Guespin, J., 2004.
\titleref{Application of formal methods to biological regulatory
networks: extending Thomas' asynchronous logical approach with
temporal logic.}
\pr{J. Theor. Biol.} \nr{229}, 339-347.
\bibitem[Bertsekas and Tsitsiklis, 1989]{tsi}
Bertsekas, D.P. \& Tsitsiklis, J.N., 1989. 
\pr{Parallel and Distributed Computation, Numerical Methods}.
Prentice Hall, New Jersey.
\bibitem[Bodnar, 1997]{b97}
Bodnar, J. W., 1997.  
\titleref{Programming the {\it Drosophila} Embryo.} 
\pr{J. Theor. Biol.} \nr{188}, 391-445.
\bibitem[Cadigan \ea, 1994]{cgg94} 
Cadigan, K. M., Grossniklaus, U. \& Gehring, W. J., 1994. 
\titleref{Localized expression of {\it sloppy paired} protein 
maintains the polarity of {\it Drosophila} parasegments.} 
\pr{ Genes \& Dev.} \nr{8}, 899-913.
\bibitem[Cadigan \& Nusse 1997]{cn97}
Cadigan, K. M., \& Nusse, R. (1997) 
\titleref{Wnt signaling: a common theme in animal
development.} \pr{Genes Dev.} {\bf 11}, 3286-3305.
\bibitem[Carlson and Doyle, 2002]{cd02}
Carlson, J. M. \& Doyle, J., 2002.
\titleref{Complexity and robustness}.
\pr{Proc. Nat. Acad. Sci.} \nr{99}, 2538-2545.
\bibitem[Conant and Wagner, 2004]{cw04}
Conant, G. C.\& Wagner A., 2004.
\titleref{Duplicate genes and robustness to transient gene knock-downs in 
Caenorhabditis elegans.}
\pr{Proc. R. Soc. Lond. B Biol. Sci.} \nr{271}, 89-96. 
\bibitem[Davidson \ea, 2002]{d02}
Davidson, E. H., {\it et al.}, 2002. 
\titleref{A genomic regulatory network for development.}
\pr{Science} \nr{295}, 1669-1678.
\bibitem[de Jong \ea, 2004]{jg04}
de Jong, H. {\it et al.}, 2004. 
\titleref{Qualitative simulation of genetic regulatory networks using 
piecewise-linear models.}
\pr{Bull. Math. Bio.} \nr{66}, 301-340.
\bibitem[DiNardo \ea, 1988]{dshk88} 
DiNardo, S., Sher, E., Heemskerk-Jongens, J., Kassis, J. A. \&  O'Farrell,
 P. H., 1988. 
\titleref{Two-tiered regulation of spatially patterned {\it engrailed} gene expression 
during {\it Drosophila} embryogenesis}. \pr{Nature} \nr{332}, 45-53.
\bibitem[Eaton \& Kornberg 1990]{ek90} 
 Eaton, S. \& Kornberg, T. B. (1990) 
 \titleref{Repression of ci-D in posterior
 compartments of Drosophila by {\it engrailed}.} \pr{Genes. Dev.} {\bf 4}, 
 1068-1077.
\bibitem[Eldar \ea, 2002]{eb02}
Eldar, A., Dorfman, R., Weiss, D., Ashe, H., Shilo, B.-Z. \& Barkai, N., 2002. 
\titleref{Robustness of the BMP morphogen gradient in {\it Drosophila} 
embryonic patterning.}
\pr{Nature} \nr{419}, 304-308.
\bibitem[Feller, 1970]{feller}
Feller, W., 1970.
\pr{An Introduction to Probability Theory and its Applications (Third edition, revised)}.
John Wiley \& Sons, New York, 1970.
\bibitem[Gallet \ea, 2000]{gakt00} 
Gallet, A., Angelats, C., Kerridge, S. \&  Th\'erond, P. P., 2000.  
\titleref{Cubitus interruptus-independent transduction of the Hedgehog signal in {\it Drosophila}},
\pr{Development} \nr{127}, 5509-5522.
\bibitem[Ghysen and Thomas, 2003]{gt03}
Ghysen, A. \& Thomas, R., 2003.
\titleref{The formation of sense organs in {\it Drosophila}: A logical approach.} 
\pr{BioEssays} \nr{25}, 802-807.
\bibitem[Glass and Kauffman, 1973]{gk73}
Glass, L. \&  Kauffman, S. A., 1973.
\titleref{The logical analysis of continuous, nonlinear biochemical control networks.} 
\pr{J. Theor. Biol.} \nr{39}, 103-129.
\bibitem[Grossniklaus \ea, 1992]{gpg92}
Grossniklaus, U., Pearson, R. K. \&  Gehring, W. J. (1992) The {\it Drosophila
sloppy paired} locus encodes two proteins involved in segmentation that show
homology with mammalian transcription factors.{\it Genes Dev.} {\bf 6}, 
1030-1051.
\bibitem[Gursky \ea, 2001]{grs01}
Gursky, V. V., Reinitz, J. \& Samsonov, A. M., 2001. 
\titleref{How gap genes make their domains: An analytical study based 
on data driven approximations.} 
\pr{Chaos} \nr{11}, 132-141.
\bibitem[Hidalgo and Ingham, 1990]{hi90} 
Hidalgo, A. \& Ingham, P., 1990. 
\titleref{Cell Patterning in the {\it Drosophila} segment:
spatial regulation of the segment polarity gene {\it patched}}. 
\pr{Development} \nr{110}, 291-301.
\bibitem[Hooper and Scott, 1992]{hs92} 
Hooper, J. E. \& Scott, M. P., 1992. 
\titleref{The Molecular Genetic Basis of Positional
Information in Insect Segments.} 
In: \pr{Early Embryonic Development of Animals} 
(ed. Hennig, W.) 1-49, Springer, Berlin.
 \bibitem[Ingham 1998]{i98}
Ingham, P. W. (1998) Transducing hedgehog: the story so far, {\it EMBO J.} 
{\bf 17}, 3505-3511.
\bibitem[Ingham \& McMahon 2001]{im01}
Ingham, P. W. \& McMahon, A. P. (2001) Hedgehog signaling in animal development:
paradigms and principles, \pr{Genes Dev.} {\bf 15}, 3059-3087.
\bibitem[Ingham \ea, 1991]{itn91}
Ingham, P.W., Taylor, A. M. \& Nakano, Y., 1991. 
\titleref{Role of the {\it Drosophila
patched} gene in positional signaling}. \pr{Nature} \nr{353}, 184-187. 
\bibitem[Jeong \ea, 2000]{jtaob00}
Jeong, H., Tombor,  B., Albert, R., Oltvai, Z.N. 
\& Barab\'asi, A.-L., 2000. 
{\titleref The large-scale organization of metabolic networks.}
\pr{Nature} \nr{407}, 651-654. 
\bibitem[Jeong \ea, 2001]{jmbo01}
Jeong, H., Mason, S., Barab\'asi, A.-L. \& Oltvai, Z. N., 2001. 
\titleref{Lethality and centrality in protein networks.}
\pr{Nature} \nr{411}, 41-42.
\bibitem[Kauffman, 1993]{k93}
Kauffman, S. A., 1993. 
\pr{The origins of Order}. Oxford University Press, New York.
\bibitem[Kauffman \ea, 2003]{kpst03}
Kauffman, S., Peterson, C., Samuelsson, B., Troein, C., 2003.
\titleref{Random Boolean network models and the yeast transcriptional network.}
\pr{Proc. Natl. Acad. Sci. USA.} \nr{100} 14796-9.
\bibitem[Martinez-Arias \ea, 1988]{mbi88} 
Martinez-Arias, A., Baker, N. \& Ingham, P. W., 1988. \titleref{Role of segment polarity
genes in the definition and maintenance of cell states in the {\it Drosophila }
embryo}. \pr{Development} \nr{103}, 157-170.
\bibitem[Mendoza \ea, 1999]{mta99}
Mendoza, L., Thieffry, D. \& Alvarez-Buylla, E. R., 1999. 
\titleref{Genetic control of flower morphogenesis in 
{\it Arabidopsis thaliana}: a logical analysis.} 
\pr{Bioinformatics} \nr{15}, 593-606.
\bibitem[Ohlmeyer \& Kalderon 1998]{ok98} 
Ohlmeyer, J. T. \& Kalderon, D. (1998) hedgehog stimulates maturation of
Cubitus interruptus into a labile transcriptional activator. {\it Nature } 
{\bf 396}, 749-753.
\bibitem[Pfeiffer \& Vincent 1999]{pv99}
Pfeiffer, S. \&  Vincent, J.-P. (1999) 
\titleref{Signaling at a distance:Transport of
Wingless in the embryonic epidermis of  {\it Drosophila}.} {\it Cell \& Dev. 
Biol.} {\bf 10}, 303-309.
\bibitem[Rao \ea, 2002]{rwa02}
Rao, C.V., Wolf, D.M. \& Arkin, A. P., 2002. 
\titleref{Control, exploitation and tolerance of intracellular noise.}
\pr{Nature} \nr{420}, 231-237.
\bibitem[Reinitz and Sharp, 1995]{rs95}
Reinitz, J. \& Sharp, D. H., 1995. 
\titleref{Mechanism of eve stripe formation.} 
\pr{Mechanisms of Development} \nr{49}, 133-158.
\bibitem[S\'anchez and Thieffry, 2001]{st01}
S\'anchez, L., \& Thieffry, D., 2001.
\titleref{A logical analysis of the {\it Drosophila} gap-gene system.} 
\pr{J. Theor. Biol.} \nr{211}, 115-141.
\bibitem[Schwartz \ea, 1995]{slnk95} 
Schwartz, C., Locke, J., Nishida, C. \&  Kornberg, T. B., 1995. \titleref{Analysis of {\it
cubitus interruptus} regulation in {\it Drosophila} embryos and imaginal disks}.
\pr{Development} \nr{121}, 1625-1635.
\bibitem[Tabata \ea, 1992]{tek92} 
Tabata, T., Eaton, S. \&  Kornberg, T. B., 1992.
\titleref{ The {\it Drosophila hedgehog}
gene is expressed specifically in posterior compartment cells and is a target of
{\it engrailed} regulation.}
\pr{Genes \& Dev.} \nr{6}, 2635-2645.
\bibitem[Taylor \ea, 1993]{tnmi93} 
Taylor, A. M., Nakano, Y., Mohler, J. \&  Ingham, P. W., 1993. 
\titleref{ Contrasting distributions of patched and hedgehog proteins in the 
{\it Drosophila } embryo.}
\pr{Mechanisms of Development }\nr{42}, 89-96.
\bibitem[Thomas, 1973]{t73}
Thomas, R., 1973. 
\titleref{Boolean formalization of genetic control circuits.} 
\pr{J. Theor. Biol.} \nr{42}, 563-585.
\bibitem[von Dassow \ea, 2000]{dmmo00} 
von Dassow, G., Meir., E., Munro, E. M., \& Odell, G. M., 2000. 
\titleref{The segment polarity network is a robust developmental
  module.} 
\pr{Nature} \nr{406}, 188-192.
\bibitem[Wolpert \ea, 1998]{w98}
Wolpert, L., Beddington, R., Brockes, J., Jessell, T., Lawrence, P., \&
Meyerowitz, E., 1998. 
\pr{Principles of Development} 
Current Biology Ltd., London.
\bibitem[Yuh \ea, 2001]{ybd01}
Yuh, C. H., Bolouri, H., Bower, J. M. and Davidson, E. H., 2001. 
\titleref{ A logical model of cis-regulatory control in a eukaryotic system}. 
In: \pr{Computational Modeling of Genetic and Biochemical Networks} 
(eds. Bower, J. M. and  Bolouri, H.), 73-100, MIT Press, Cambridge, MA.

\end{thebibliography}
\end{document}